\documentclass[useAMS,usegraphicx,usenatbib]{mn2e}
\usepackage{aas_macros}     

\newcommand{\mc}[3]{\multicolumn{#1}{#2}{#3}}

\title[The detectability of BAO in future galaxy surveys]
{The detectability of baryonic acoustic oscillations in future galaxy surveys}
\author[Angulo  et al.]{
\parbox[h]{160mm}{
R. E. Angulo\thanks{E-mail: raul.angulo@durham.ac.uk},
C. M. Baugh\thanks{E-mail: c.m.baugh@durham.ac.uk},
C. S. Frenk\thanks{E-mail: c.s.frenk@durham.ac.uk},
C. G. Lacey\thanks{E-mail: cedric.lacey@durham.ac.uk}.
}\vspace{6pt}\\
Institute for Computational Cosmology, Department of Physics, 
University of Durham, South Road, Durham, DH1 3LE, UK. 
\vspace*{-0.5cm}}

\begin{document}
\date{\today}
\pagerange{\pageref{firstpage}--\pageref{lastpage}} \pubyear{2007}
\maketitle
\label{firstpage}

\begin{abstract}
We assess the detectability of baryonic acoustic oscillations (BAO) 
in the power spectrum of galaxies using ultra large volume N-body 
simulations of the hierarchical clustering of dark matter 
and semi-analytical modelling of galaxy formation. 
A step-by-step illustration is given of the various effects 
(nonlinear fluctuation growth, peculiar motions, nonlinear 
and scale dependent bias) which systematically change the form 
of the galaxy power spectrum on large scales from the simple 
prediction of linear perturbation theory. Using a new method to 
extract the scale of the oscillations, we nevertheless find that 
the BAO approach gives an unbiased estimate of the sound horizon 
scale. Sampling variance remains the dominant source of error 
despite the huge volume of our simulation box ($=2.41 h^{-3}{\rm Gpc}^{3}$). 
We use our results to forecast the accuracy with which forthcoming 
surveys will be able to measure the sound horizon scale, $s$, 
and, hence constrain the dark energy equation of state parameter, $w$ (with 
simplifying assumptions and without marginalizing over the other cosmological 
parameters). Pan-STARRS could potentially yield a measurement 
with an accuracy of $\Delta s/s = 0.5-0.7 \% $ (corresponding to 
$\Delta w \approx 2-3\% $), which is competitive with the proposed 
WFMOS survey ($\Delta s/s = 1\% $ \,\, $\Delta w \approx 4 \% $). 
Achieving $\Delta w \le  1\% $ using BAO alone is beyond any 
currently commissioned project and will require an all-sky 
spectroscopic survey, such as would be undertaken by the SPACE 
mission concept under proposal to ESA.   
\end{abstract}


\section{Introduction}

The discovery that the rate of expansion of the Universe is apparently 
accelerating was one of the key advances in physical cosmology in 
the 1990s (\citealt{Riess1998, Perlmutter1999}). Understanding 
the nature of the dynamically dominant dark energy, which is believed 
to be responsible for this behaviour, is one of the biggest 
challenges now facing cosmologists. 

Over the past decade our knowledge of the basic cosmological parameters, 
which describe the content of the Universe, its expansion history and ultimate 
fate has improved tremendously. This progress is the result of advances 
on two fronts: the advent of datasets which have provided fresh  
views of the Universe with unprecedented detail and the development of the 
theoretical machinery required to interpret these new measurements. 
Currently, the values of many cosmological parameters are known to an accuracy 
of around $10\%$ (albeit with caveats regarding degeneracies between 
certain combinations of parameters and also regarding the precise number of parameters 
that are allowed to vary in the cosmological model; see, for example,
\citealt{Sanchez2006}).  

The cold dark matter (CDM) model has emerged as the most plausible description
of our Universe. In the most successful version of this model, more than $70\%$
of the density required to close the Universe is in the form of dark
energy. Currently, there is no model which can reconcile the magnitude of the
dark energy component with the value expected from particle physics arguments. A
simple phenomenological description of the dark energy is provided by the
equation of state that relates its pressure, $P$, and density, $\rho$, which is
encapsulated in the parameter $ w = P/\rho c^{2}$. If the dark energy has the
form of the cosmological constant, $w=-1$. The indications are that the dark
energy now has a form close to that expected for a cosmological constant
(\citealt{Riess2004, Sanchez2006}). However, in the absence of a theoretical
model for the dark energy, it is possible that the equation of state could
depend on space and/or time.

A whole range of experiments and surveys is being planned which number amongst
their goals determining the equation of state of the dark energy as a function
of redshift (for a discussion, see \citealt{Albrecht2006} and 
\citealt{Peacock2006}). Several techniques are being considered,
which are sensitive to the influence of the dark energy on various features of 
the cosmological world model. These include the Hubble diagram of Type IA 
supernovae, counts of clusters of galaxies, the weak gravitational lensing pattern
of faint galaxies and the measurement of the baryonic acoustic oscillation scale
in the matter distribution as a function of redshift. The measurements and data analysis
required to obtain useful constraints on the equation of state parameter are so
demanding, and so open to potential systematic errors, that it is necessary to
pursue as many different avenues as possible.

In this paper, we focus on the test using the baryonic acoustic oscillations
(BAO). The BAO is the name given to a series of peaks and troughs on scales on
the order of $100 \,h^{-1}\,$Mpc, imprinted on the power spectrum of matter
fluctuations prior to the epoch of last scattering, when the matter and
radiation components of the Universe were coupled (\citealt{PY1970}). 
The BAO are the counterpart of the acoustic peaks seen in the power spectrum of 
the temperature of the cosmic microwave background radiation, though they have a
different phase and a much smaller amplitude (\citealt{SZ1970, 
Press1980, Hu1996, EH1998,Meiksin1999}). The wavelength of the BAO 
is related to the size of the sound
horizon at recombination. This does not depend on the amount or nature of the
dark energy, but on the physical density of matter ($\Omega_m h^2$) and 
baryons ($\Omega_b h^2$).
Given the values of these parameters, for example, from the cosmic
microwave background or large scale structure data, the sound horizon scale is
known and can be treated as a standard ruler. The {\it apparent} size of this
feature in the power spectrum of galaxies or galaxy clusters does depend on the
dark energy and its equation of state through the angular diameter distance-redshift relation 
(e.g. \citealt{BG2003, HuHaiman2003})

BAO in the galaxy distribution were first glimpsed in
the early stages of the ``2-degree-field galaxy redshift survey''
(\citealt{Percival2001}) and finally detected in the power spectrum of the 
completed 2dFGRS (\citealt{Cole2005}). The equivalent feature, a spike, was 
also found in the correlation function measured from the luminous red galaxy (LRG)
sample of the Sloan Digital Sky Survey (SDSS) (\citealt{Eisenstein2005}). 
Cole et al used the BAO to constrain the parameter combination 
($\Omega_{\rm M}/\Omega_{\rm b}$, $\Omega_{\rm M}$) (where $\Omega_{\rm M}$ 
and $\Omega_{\rm b}$ denote the matter and baryon density parameters 
respectively). Eisenstein et~al used the location of the spike in the 
correlation function to constrain the absolute distance to the median 
redshift of the SDSS LRG sample and hence constrained the value of 
$\Omega_{\rm M}$. H\"{u}tsi (2006a,b) carried out a power spectrum analysis 
of a similar LRG sample, and combined this measurement with other datasets 
to constrain the values of cosmological parameters. More recently, the 
BAO have been extracted from the power spectrum measured from a much 
larger sample of SDSS LRGs to constrain $\Omega_{\rm M}$ 
and $\Omega_{\rm b}/\Omega_{\rm m}$ ( \citealt{Tegmark2006, Blake2007, Padmanabhan2007, 
Percival2007}). To date, measurements of the 
BAO have only yielded constraints on the dark energy equation of state 
when combined with other datasets, such as the spectrum of temperature 
fluctuations in the microwave background or when restrictive priors have 
been adopted on certain parameters, such as the Hubble constant.

The bulk of the work in the literature on the usefulness of the BAO has relied
upon linear perturbation theory to assess the detectability of the features and to
forecast the errors on the recovered value of $w$ (\citealt{BG2003, HuHaiman2003, GB2005,
BlakeBridle2005, Blake2006, Parkinson2007}). There are, however, a range of dynamical and
statistical effects which can alter the appearance of the power spectrum
relative to the linear theory prediction, even on the scale of the BAO, which we
review in this paper (\citealt{Seo2003, Angulo2005, Millennium, Seo2005,
Eisenstein2006}). Some simulation work has been done to study these effects, mostly
using computational cubes of side $500 \,h^{-1}\,$Mpc (\citealt{Seo2003, Seo2005,
Millennium, Eisenstein2006}). These are only a small factor (2-3) bigger than the
scale of the fluctuations of interest. Calculations with small boxes are subject 
to large sampling fluctuations and may even miss some features of the nonlinear 
growth of large scale fluctuations through the absence of long wavelength density 
fluctuations (\citealt{Crocce2006b}).
Very recently, larger simulation volumes have been used, of around
a cubic gigaparsec and larger (\citealt{Schulz2006, Huff2006, Angulo2005,
Koehler2006}). However, such studies have tended to have relatively poor mass
resolution, making it difficult to model galaxies without resorting to
simplified biasing prescriptions (e.g. \citealt{Cole1998}).

Given the significant commitment of resources required by the proposed galaxy
surveys and the level of precision demanded by the BAO approach, it is
imperative to ensure that accurate theoretical predictions are available both to
help in the design of the survey strategy and to extract the maximum amount of
information from the observations. This is a tough challenge computationally,
because it requires ultra-large volume N-body simulations with sufficient mass
resolution to identify the haloes likely to host the galaxies to be seen in the
surveys, and a realistic model to populate these haloes with galaxies.

In this paper, we use a combination of suitable N-body simulations and a
semi-analytical model of galaxy formation to assess the visibility of the
BAO. In Section 2, we describe the suite of N-body simulations used and outline
the semi-analytical model. Section 3 gives a blow-by-blow account of how the
power spectrum changes relative to the simple prediction of linear perturbation
theory, as additional layers of realism are added to the modelling, starting
with dark matter and ending with galaxies. We set out our approach for
constraining the dark energy equation of state in Section 4, and present our
results in Section 5. We give our conclusions in Section 6.

\section{Method}

In this section, we introduce the theoretical tools used to produce synthetic
galaxy catalogues.  First, we describe the N-body simulations (\S 2.1) which
consist of a high resolution run (\S 2.1.1) and an ensemble of lower resolution
runs (\S 2.1.2).  Next, we discuss the measurement of power spectra from
discrete distributions of objects and use the ensemble of low resolution
simulations to estimate the errors on the power spectrum measurement (\S
2.1.3). In the second part of this section, we explain how a galaxy formation
model is used to populate the high resolution N-body simulation with
galaxies (\S 2.2).

\begin{table}
\begin{center}
\begin{tabular}{rrrrrrrrr}
\hline
\hline
&  $N_{\rm p}$ & $m_{\rm dm}$         & $\epsilon$     \\ 
&              & [$h^{-1}~$M$_\odot$] & [$h^{-1}$~kpc] \\
\hline
\textsc{BASICC}  & $3.03\times10^{9}$ & $5.49\times10^{10}$ & $50$ \\
\textsc{L-BASICC}   & $8.99\times10^{7}$ & $1.85\times10^{12}$ & $200$ \\
\hline
\end{tabular}
\end{center}
\caption{The values of some of the basic parameters used in the simulations. 
The columns are as follows: (1) The name of the simulation. (2) The number of 
particles. (3) The mass of a dark 
matter particle. (4) The softening parameter used in the gravitational force. 
In both cases, the length of the computational box is $1340 \,h^{-1}\, {\rm Mpc}$, 
and the same cosmological parameters are used, as given in Section 2.1.
}
\label{tab:params}
\end{table}

\subsection{N-Body Simulations}
\begin{figure}
\includegraphics[width=8.5cm]{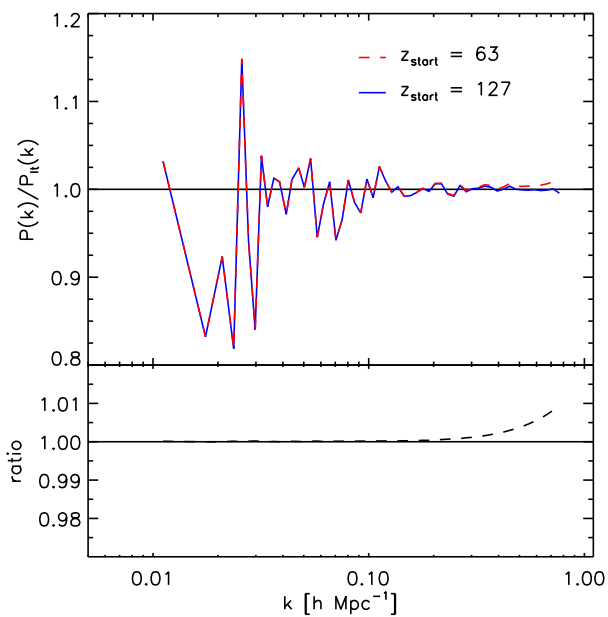}
\caption{
A test of the choice of starting redshift used in the N-body simulations. 
The upper panel compares the power spectrum measured at $z=15$ in the {\tt BASICC} 
when the simulation is started at $z=63$ (dashed red curve) and at $z=127$ 
(solid blue curve). The power spectra plotted in the upper panel have been 
divided by the linear perturbation theory prediction for the dark matter 
power spectrum at $z=15$. The lower panel shows the ratio between the  
power spectrum measured from the simulation started at redshift 63 to  
that measured from the run which started at redshift 127.
\label{fig:trans}
}
\end{figure}

The N-body method is a long-established computational technique which is used to
follow the growth of cosmological structures through gravitational instability
(see, for example, the reviews by Bertschinger 1998 and Springel, Frenk \&
White 2006). Our goal in this paper is to simulate the formation of structure
within a sufficiently large volume to follow the growth of fluctuations
accurately on the scale of the BAO, and with similar statistics
for power spectrum measurements to those expected in forthcoming surveys. At the same
time, we require a mass resolution which is adequate to identify the dark matter
haloes likely to host the galaxies which will be seen in these surveys. To
achieve these aims, we use a memory-efficient version of the {\tt GADGET-2} code
of \cite{Springel2005}, which was kindly provided to us by Volker Springel and
the Virgo Consortium.

We use two types of calculation: a high resolution simulation, labelled the
``Baryon Acoustic Simulation at the ICC' or {\tt BASICC}, which is able to track
galactic haloes, and an ensemble of lower resolution simulations, labelled
{\tt L-BASICC}, which we use to study the statistics of power spectrum measurements on
large scales. Here, we describe some of the common features of the simulations,
before moving on to outline specific details in \S 2.1.1 and \S 2.1.2.

We adopt a $\Lambda$CDM cosmology with the same parameters used in the
Millennium Simulation (\citealt{Millennium}), which are broadly consistent with
the latest constraints from the cosmic microwave background data and large scale
structure measurements (\citealt{Sanchez2006, Spergel2007}). The values of the
parameters are: the matter density parameter, $\Omega_{\rm M}=0.25$, the energy
density parameter for the cosmological constant, $\Omega_{\Lambda}=0.75$, the
normalization of density fluctuations, $\sigma_{8} = 0.9$ and Hubble constant,
$h=H_{0}/(100{\rm km s}^{-1}{\rm Mpc}^{-1})=0.73$.

Due to memory restrictions, the Fourier mesh used to set up the initial particle
displacements has a dimension of $1580^3$ grid points which is not commensurate
with the cube root of the particle number mesh. We therefore avoided using a
regular particle grid to set up the initial conditions, as this would have led
to a spurious feature in the power spectrum of the initial conditions at the
beat frequency between the particle grid and the Fourier mesh. Instead, we used
a glass-like distribution (\citealt{White1994, Baugh1995}). The input power 
spectrum of density fluctuations in linear perturbation theory is calculated using 
the {\tt CAMB} package of
\cite{Lewis2000}. The amplitude of the Fourier modes is drawn from a Rayliegh
distribution with mean equal to the linear theory power spectrum and the phase
is drawn at random from the interval $0$ to $2\pi$. The initial density field is
generated by perturbing particles from the glass-like distribution, using 
the approximation of Zel'dovich (1970).

The simulations were started at a redshift of $z=63$. The Zel'dovich (1970)
approximation used to set up the initial pattern of density fluctuations
produces transients which can be seen in clustering signal measured for the dark
matter at expansion factors close to the starting redshift
(\citealt{Efstathiou1985,Baugh1995,Crocce2006}). Later on, we will use the power
spectrum from a high redshift output from the simulation, $z=15$, as a proxy for
linear perturbation theory, so it is important to check that this power spectrum
in particular, and also the power spectra measured at all subsequent outputs are
insensitive to the choice of starting redshift. We test this by comparing the
power spectrum of the dark matter at $z=15$ in our standard run with the
spectrum measured in a test run which started at $z=127$, but which did not run
all the way through to $z=0$. The top panel of Fig.~\ref{fig:trans} shows that
the power spectra measured for the dark matter in these two cases, divided by
the power spectrum predicted by linear perturbation theory at $z=15$. The
fluctuations in the measured power at low wavenumbers around the linear theory 
prediction reflect the sample
variance noise which is not negligible even in a simulation of the volume of the
{\tt BASICC}. The lower panel in Fig.~\ref{fig:trans} shows the $z=15$ power spectrum
measured from the run started at $z=63$ divided by that measured from the run
started at $z=127$. At large wavenumbers, the effect of transients is visible, 
although quite small, $\sim 1\%$. The focus of this paper, however, is the form
of the power spectrum over wavenumbers smaller than $k=0.4\,h {\rm Mpc}^{-1}$,
for which the spectra measured at $z=15$ for the two different choices of
starting redshift agree to better than $0.3\%$. Our results are therefore
unaffected by any transients resulting from the use of the Zel'dovich
approximation.

\subsubsection{The high resolution simulation: the {\tt BASICC}}

The {\tt BASICC} simulation covers a comoving cubical region of side 
$1340\,h^{-1}\,$Mpc, in which the dark matter is represented by more
than 3 billion ($1448^3$) particles. The equivalent Plummer softening length in
the gravitational force is $\epsilon = 50\,h^{-1}\,\rm{kpc}$, giving a dynamic
range in length of almost 27,000. The volume of the computational box,
$2.41\,h^{-3}\,{\rm Gpc}^{3}$, is almost twenty times the volume of the Millennium
Simulation (\citealt{Millennium}), and more than three times the volume of the
catalogue of luminous red galaxies from the SDSS used to detect the acoustic
peak by \cite{Eisenstein2005}. The {\tt BASICC} volume is within a factor of two of
that proposed for a survey with WFMOS at $z \sim 1$ (\citealt{GB2005}). The
simulation occupied the full 0.5 Terabytes of RAM of the second upgrade of the
Cosmology Machine at Durham. The run took 11 CPU days on 506 processors, the
equivalent of 130,000 cpu-hours. 

The particle mass in the {\tt BASICC} simulation is $m_{\rm p} = 5.49\times
10^{10}\,h^{-1}\,M_{\odot}$. This is approximately 64 times larger than the
particle mass used in the Millennium Simulation. The mass resolution limits the
usefulness of dark matter halo merger trees from the {\tt BASICC}, so we have chosen
to output at a modest selection of redshifts: z=0, 0.3, 0.5, 1, 2, 3, 4, 6, 8,
10, 15 and 63. Each of these outputs occupies $\sim100\,$ {\tt Gb} of disk
space.
:
In each snapshot we have identified groups of dark matter particles using a
friends-of-friends algorithm (\citealt{Davis1985}) with a linking length of
$0.2$ times the mean inter-particle separation. We have stored groups with 10 or
more particles, i.e. haloes more massive than $5.49 \times
10^{11}\,h^{-1}\,\rm{M_{\sun}}$. There are 17\,258\,579 haloes in the $z=0$
output of the simulation with ten or more particles. The most massive halo has a
mass of $6.74 \times 10^{15}\,h^{-1}\,M_{\odot}$ and $860$ haloes have a mass in
excess of the Coma cluster ($\approx 10^{15}\,h^{-1}\,M_{\odot}$).

The {\tt BASICC} simulation sits between the Millennium and Hubble Volume
(\citealt{Evrard2002}) simulations. Its unique combination of mass resolution
and volume makes it ideal for studying the large scale distribution of galaxies
and clusters alike.

\subsubsection{The ensemble of low resolution simulations: {\tt L-BASICC}}

We also generated an ensemble of 50 ``low-resolution'' simulations to study the
sample variance in the {\tt BASICC} and to test an analytic model for the errors
expected on measurements of the power spectrum, which we discuss in the next
subsection. These low resolution runs ({\tt L-BASICC}) have exactly the same
cosmological parameters as the {\tt BASICC} and the same box size (see Table 1), but
they have fewer particles ($448^3$). For each realization, a different random
seed is used to set up the initial density field. The starting redshift of these
simulations is $z=63$. The particle mass is comparable to
that employed in the Hubble Volume simulation (\citealt{Evrard2002}). Each {\tt L-BASICC}
simulation took $0.8$ days to run on $16$ processors of the third
upgrade of the Cosmology Machine. The total volume of the ensemble is $120\,
h^{-3}\,{\rm Gpc}^{3}$, more than four times that of the Hubble Volume, making
this a unique resource for studying the frequency of rare objects in a
$\Lambda$CDM universe. For {\tt L-BASICC}, the position and
velocity are stored for every particle at 4 output times (z = 0.0, 0.5, 0.9,
3.8); we also produce a halo catalogue at each redshift retaining objects with
ten or more particles (corresponding to a mass of
$1.8 \times 10^{13}\,h^{-1}\,\rm{M_{\sun}}$). As we shall see in later sections, the
ensemble allows us to assess whether or not a particular result is robust or
simply due to sampling fluctuations. Due to their limited mass resolution, it is
not feasible to populate these simulations with galaxies using the method
outlined below (\S 2.2).

\subsubsection{Power spectrum estimation and errors}

\begin{figure}
\includegraphics[width=8.5cm]{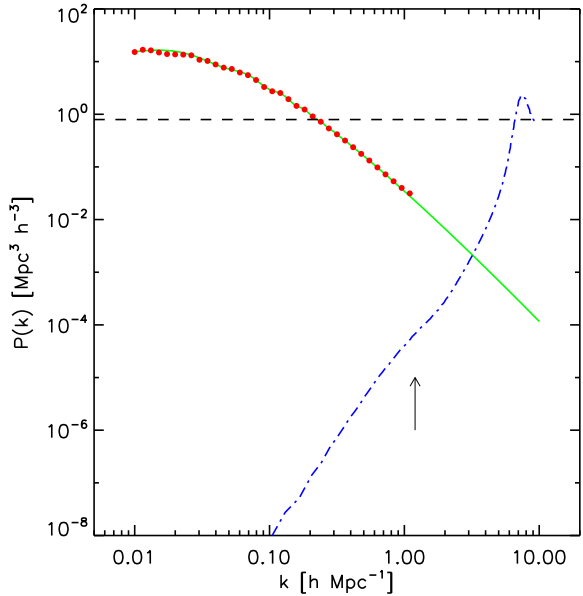}
\caption{
The power spectrum of the dark matter in real-space measured at the 
starting redshift of the {\tt BASICC}, $z=63$ (red points). The 
corresponding prediction of linear perturbation theory is shown by the 
green (solid) line. The blue (dot-dashed) curve shows the power 
spectrum of the {\it unperturbed} glass-like distribution of particle positions. 
The dashed line shows the Poisson noise expected for the number density of 
dark matter particles used in the {\tt BASICC}. 
The noise of the initial particle distribution is much less than Poisson. 
The arrow marks the position of the Nyquist 
frequency of the FFT grid. 
\label{fig:glass}
}
\end{figure}

\begin{figure}
\includegraphics[width=8.5cm]{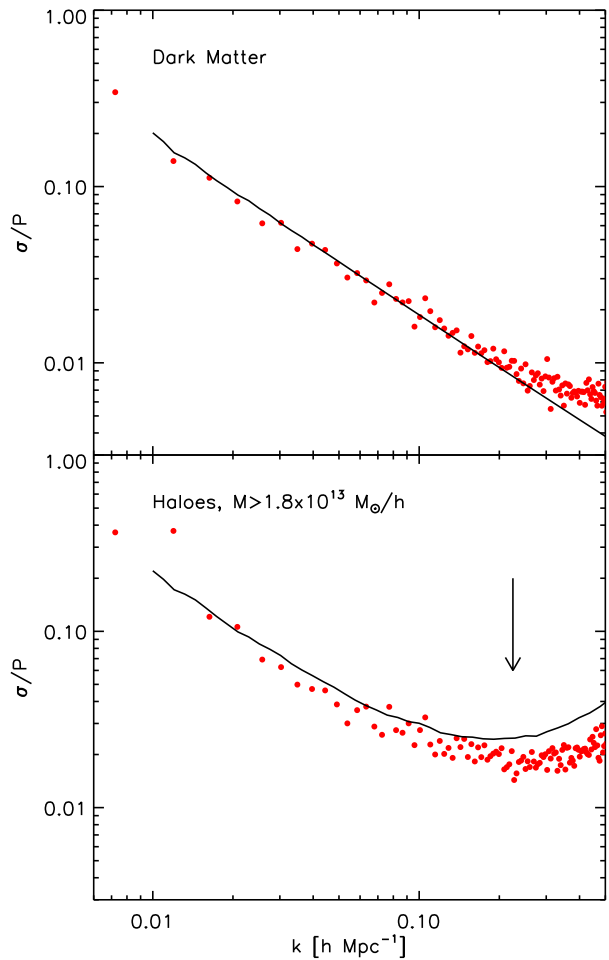}
\caption{
The fractional error in the power spectrum of the dark matter (top panel) 
and in the power spectrum of haloes more massive than $1.8 \times 10^{13}\, 
h^{-1}\,M_{\odot}$ (bottom panel), estimated using the low resolution 
simulations from the dispersion of $P(k)$ around the ensemble mean. 
The smooth black curves show the error predicted by the analytical 
expression given in Eq.~\ref{eq:err}. The red points show the scatter from 
the ensemble of low resolution simulations. The arrow in the bottom panel 
shows the wavenumber for which $\bar{n}P(k=0.2 h {\rm Mpc}^{-1})=1$. 
\label{fig:err}}
\end{figure}

The two point statistics of clustering, the correlation function, and its
Fourier transform, the power spectrum, $P(k)$, are the most commonly employed
measurements of clustering. In this paper we focus on the power spectrum; in
Sanchez et~al. (2007, in preparation), we address the visibility of the acoustic
oscillations in the correlation function. The standard way to quantify the amplitude
of a density fluctuation is by means of the density contrast, $\delta (x,t) =
(\rho(x,t)-{\bar{\rho}})/\bar{\rho}$. If we consider the Fourier transform of
the density contrast, $\rho_{\rm k}$, then the power spectrum is defined as the
modulus squared of the mode amplitude, $P(k) = \langle | \delta_{\rm k}) |^{2}
\rangle$.

There are two steps in the computation of the power spectrum from a distribution
of discrete objects, such as dark matter particles, dark haloes or
galaxies. Firstly, a density field is constructed by assigning the objects to
mesh points on a cubic grid. In the simplest mass assignment scheme, the nearest grid
point, the contribution of each object to the density field is confined to the
cell in which it is located. In higher-order assignment schemes, the mass of the
particle is shared with adjacent cells. Here, we use the cloud-in-cell
assignment scheme (see \citealt{Hockney1981}). Secondly, we perform a Fast
Fourier Transform of the density field. The power spectrum is obtained by
spherically averaging the resulting Fourier mode amplitudes in annuli of radius
$\delta k = 2 \pi / L = 0.0047 \,h\,{\rm Mpc}^{-1}$.

The mesh we use to store the density field has $N^{3}_{\rm FFT} = 512^3$ grid
points. Estimating the density on a grid alters the form of the power spectrum
at wavenumbers approaching the Nyquist frequency of the grid ($k_{\rm Nyquist} =
2 \pi/ L \,\, N_{\rm FFT}/ 2 = 1.2 h {\rm Mpc}^{-1}$ in our case). The degree of
modification and the precise wavenumber above which the power spectrum is
distorted depend upon the choice of assignment scheme (Hatton 1999;
\citealt{Jing2005}). In practice, for the size of FFT mesh we use, this has
little impact on the recovered power spectrum for wavenumbers of interest; the
measured amplitude differs by less than 1\% from the true value at a wavenumber
$k \sim 0.8 \,h\, {\rm Mpc}^{-1}$; 
in most cases we focus on the form of the power spectrum
on large scales, $k < 0.4 \,h\,{\rm Mpc}^{-1}$.  
Nevertheless,
we correct for the effects of the cloud-in-cell assignment scheme by dividing each mode 
by the Fourier transform of a cubical top hat:

\begin{equation}
\delta(k_{x},k_{y},k{z}) \Rightarrow \frac{ \delta(k_{x},k_{y},k{z}) }{ 
\textrm{{sinc}}( \frac{k_x L}{2 N_{\rm FFT}} ) \, \textrm{{sinc}}( \frac{k_y L}{2 N_{\rm FFT}} ) \, \textrm{{sinc}}( \frac{k_z L}{2 N_{\rm FFT}} ) },
\end{equation}

\noindent where

\begin{equation}
 \textrm{{sinc}}( x ) = \frac{\sin(x)}{x}.
\end{equation}
Note this is different from the approach taken by \cite{Jing2005}, who applied a
correction to the spherically averaged power spectrum.

A further possible distortion to the form of the measured power spectrum is
discreteness noise and the associated Poisson or shot noise.  Poisson-sampling a
continuous density field with point objects of space density, $\bar{n}$,
introduces a spurious contribution that should be subtracted from the measured
power spectrum: $P_{\rm corr}(k) = P_{\rm meas}(k) - 1/\bar{n}$.  In the case of dark matter
halo centres or galaxies, the need for such a correction is justified. However,
in the case of dark matter particles in our simulations, one should {\it not}
subtract Poisson shot noise from the power spectrum because the particles were
initially laid down by perturbing a glass-like configuration which is
sub-Poissonian in nature. This is clear from Fig.~\ref{fig:glass}, which shows
the power spectrum measured for the dark matter in the initial conditions of the
{\tt BASICC}. The red curve shows the spectrum measured in the simulation and the
smooth green curve shows the input spectrum predicted by linear perturbation
theory. The two agree remarkably well over a wide range of wavenumbers. The
power spectrum of the {\it unperturbed} glass-like particle distribution is
shown by the blue curve. For the wavenumbers of interest, the power spectrum of
the glass is many orders of magnitude below the discreteness noise expected for
a Poisson distribution of objects with the same space density as the dark matter
particles, as shown by the dashed line. In this paper, we do not apply any shot
noise correction to power spectra measured for the dark matter, but we do make
such a correction for spectra estimated for samples of haloes and galaxies.

To close this subsection, we turn our attention to the error on the
measurement of the power spectrum. A commonly used expression for the
fractional error in the measured power spectrum was derived by
Feldman, Kaiser \& Peacok(1994) (see also Efstathiou 1988, for a
similar argument applied to the two point correlation function):
\begin{equation}\label{eq:err}
 \frac{\sigma}{P} = \sqrt{ \frac{2}{n_{\rm modes}}} \left( 1 + \frac{1}{P \bar{n}} \right), 
\end{equation}
where $n_{\rm modes}$ is the number of Fourier modes present in a 
spherical shell of width $\delta k $, which depends upon the 
survey volume $V$: for $k \gg 2 \pi /V^{1/3}$, this is given by 
$n_{\rm modes} = V 4 \pi k^{2}{\delta }k/\left(2 \pi\right)^{3}$ . 
The first term on the right hand side of Eq.~\ref{eq:err}
quantifies the sample variance in the measurement, which decreases as
the square root of the number of modes or, equivalently, as the square
root of the volume probed. The second term arises from the
discreteness of the objects under consideration. The combination ${P
\bar{n}}$ quantifies the amplitude of the power spectrum in units of
the Poisson shot noise, effectively giving the contrast of the power
spectrum signal relative to the shot noise level.  In the case where
$P \bar{n} \gg 1$, $\sigma/P \propto 1/k $. On the other hand, when
the amplitude of the power spectrum is comparable to the shot noise,
and if $P(k) \propto k^{-1}$, then the fractional error in the power
is approximately independent of wavenumber. We have tested this
prescription in both regimes against the diagonal element of the covariance 
between power spectrum measurements extracted from the ensemble of low 
resolution simulations, as shown in
Fig.~\ref{fig:err}. Over the wavenumber range of interest, the agreement 
is reasonably good for samples in which the shot noise is negligible compared 
to the clustering signal. For samples with low contrast power measurements, 
such as is the case for dark matter haloes used in the bottom panel 
of Fig.~\ref{fig:err}, the analytic expression works well until $k \sim 0.1 h {\rm Mpc}^{-1}$ and 
then overpredicts the errors by up to 50\%.
We note that nonlinearities and the impact of the window function 
of a realistic survey could introduce off-diagonal terms in the power spectrum 
covariance matrix. In Section 5.3, we compare the constraints on the recovered 
oscillation scale using the scatter from the ensemble and using the simple 
mode-counting argument outlined above. We find good agreement which suggests 
that mode-coupling does not make a significant contributions to the errors on 
the scales relevant to the BAO.

\subsection{Modelling the formation and evolution of galaxies}
\label{ssec:galform}

\begin{figure}
  \includegraphics[width=14.5cm]{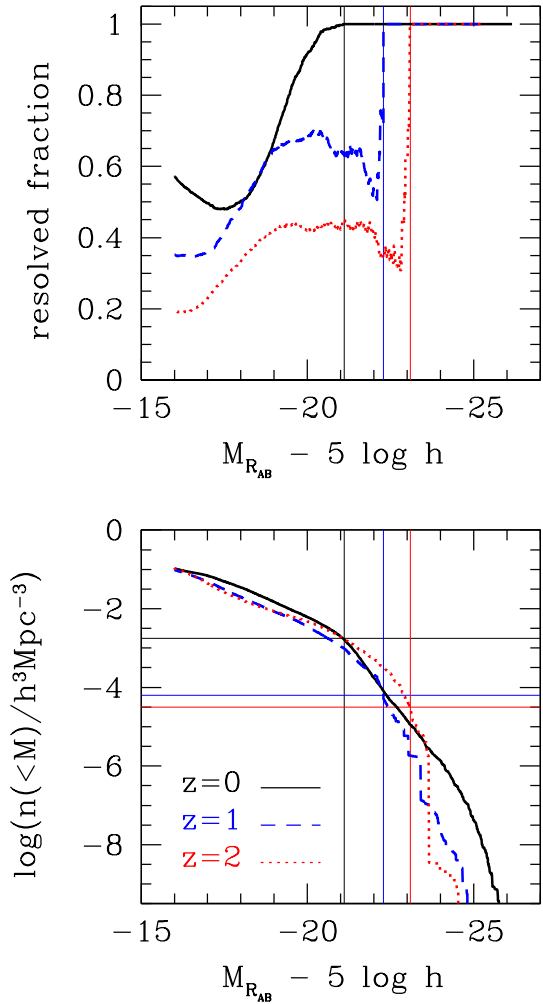}
\caption{
Upper panel: the fraction of `resolved galaxies' in the high resolution
N-body simulation as a function of magnitude, at different output
redshifts (as given by the key in the lower panel). The magnitude
is in the observer-frame $R$-band; to obtain an apparent
$R$-band magnitude, the distance modulus corresponding to the redshift
should be added to the plotted magnitude. The vertical lines mark the
magnitude at which the galaxy sample is 100\% complete at each
redshift.  Lower panel: the cumulative luminosity function of galaxies
brighter than a given $R$-band magnitude, for different redshifts as
given in the key. The vertical lines show the 100\% completeness
limits at each redshift and the horizontal lines indicate the
associated space density of galaxies.
\label{fig:comp}
}
\end{figure}

The N-body simulations described in the previous section follow the
growth of fluctuations in the mass which is dominated by collisionless
matter. To connect the predictions of the cold dark matter theory to
forthcoming galaxy surveys, we need to predict which structures host
galaxies and how galaxy properties depend on halo mass. 

Some authors have chosen to incorporate galaxies into an N-body simulation 
empirically by using a parametric model called
a halo occupation distribution function (HOD) to describe the
probability distribution of galaxies expected in haloes of a given
mass (\citealt{Benson2000}). The form of the HOD is constrained to
reproduce a particular clustering measurement, such as the galaxy correlation
function (e.g. \citealt{Peacock2000, Seljak2000, Scoccimarro2001,
Cooray2002}).  This approach has been applied to the
study of the detectability of acoustic oscillations by several authors
(\citealt{ Seo2005, Schulz2006, Huff2006}). Two assumptions are made 
when using the HOD to populate an N-body
simulation with galaxies. Firstly, the parameterization used for the
HOD is assumed to provide an accurate description of the manner in
which galaxies populate haloes across a wide range of halo
mass. Detailed comparisons between the clustering predictions made
using HODs and those obtained directly from simulations of galaxy
formation show that in practice, the HODs do a reasonable job
(\citealt{Berlind2003, Zheng2005}). Recently, one of the fundamental
assumptions which underpins the HOD approach has been called into
question. Using the Millennium simulations, \cite{Gao2005}
demonstrated that the clustering of dark matter haloes depends 
on a second parameter, such as the formation time of the halo,
in addition to halo mass (see also \citealt{Harker2006} and
\citealt{Wechsler2006}, \citealt{Wetzel2007}). In practice, for typical galaxy samples, this
effect is largely washed out due to the mix of halo properties sampled
(\citealt{Croton2007}). The second implicit assumption in the HOD
method when applied to an N-body simulation is that all of the haloes
in which galaxies are expected to be found can be resolved in the
simulation; if the mass resolution of the simulation turns out to be
inadequate, then the HOD realized will be distorted to compensate,
compared with the true, underlying HOD in the Universe.

In this paper, we take a more physical approach and make an {\it ab
initio} prediction of which dark matter haloes should contain galaxies
by modelling the physics of the baryonic component of the universe. We
do this using a semi-analytic model of galaxy formation (for a review
of this technique see \citealt{Baugh2006}). The semi-analytic model
describes the key physical processes which are thought to determine
the formation and evolution of galaxies. We use the {\tt GALFORM} code
introduced by \cite{Cole2000} and developed in a series of papers
(\citealt{Benson2002, Benson2003, Baugh2005, Bower2006}).  The
specific model we use is the one proposed by \cite{Baugh2005}, which
reproduces the abundance of Lyman-break galaxies at $z=3$ and $z=4$,
the number counts of sub-mm detected galaxies (with a median redshift
$z \sim 2$), and a rough match to the abundance of luminous red
galaxies (Almeida et~al. 2007, in preparation), whilst at the same
time giving a reasonable match to the observed properties of local
galaxies (e.g. \citealt{Nagashima2005b, Nagashima2005a, Almeida2007}).

A key advantage of using a semi-analytic model is that we can
investigate how the manner in which galaxies are selected affects the
accuracy with which the acoustic oscillations can be measured. The
model predicts the star formation history of each galaxy and uses this
to compute a spectrum, broadband magnitudes and emission line
strengths (for examples of the latter, see \citealt{leDelliou2005,
leDelliou2006}). We can therefore select samples of model galaxies by
applying precisely the same criteria which will be applied in the
proposed surveys.

Our methodology mirrors the hybrid schemes introduced by
\cite{Kauffmann1997} and \cite{Benson2000}. We use a Monte Carlo
technique to generate merger
trees for dark mater haloes since our simulation outputs do not have the
resolution in time or mass necessary to allow the construction of
merger trees. (See \citealt{Baugh2006} for a discussion of the relative
merits of these two approaches.)

We first construct a grid of halo masses at the redshift of interest,
which extends to lower mass haloes than can be resolved in the
simulation. We then generate a number of Monte-Carlo realizations of
mass assembly histories for each mass on the grid, using the algorithm
introduced by \cite{Cole2000}. The number of realizations is chosen to
allow robust predictions to be made for observables such as the galaxy
luminosity function. The halo merger history is input into the
semi-analytic code and the properties of the galaxy population are
output at the redshift for which the galaxy catalogue is to be
constructed. In the calculations in this paper, we output the
broadband magnitudes in the $R$, $I$ and $K$ bands and the equivalent
widths of $H_{\alpha}$ and OII[3727] for each galaxy. Finally, haloes
from the grid are matched with haloes of similar mass identified in
the N-body simulation. The central galaxy in each halo is assigned to
the centre of mass of the matched halo in the simulation. The
satellite galaxies are assigned randomly to dark matter particles in
the halo. Galaxies placed in the simulation box in this way are called
`resolved galaxies'.  The Monte Carlo merger trees will not, of course,
correspond in detail with those of the matched halos in the N-body
simulation. However, to the extent that the halo assembly bias
discussed by \cite{Gao2005} can be neglected, the properties of the
trees are statistically similar for haloes in the same mass range.

Because of the finite mass resolution of the N-body simulation, galaxy
samples generated by populating resolved haloes will be incomplete
fainter than some magnitude limit. In principle, since we are using
Monte-Carlo merger trees, we can follow galaxies down to arbitrarily
faint magnitudes {\it within} a resolved dark matter halo. However, as
we consider progressively fainter objects, some fraction of these
galaxies should also appear in haloes which the simulation cannot
resolve, causing the sample to become incomplete. Thus, in some
instances we need to consider galaxies which we would expect to find
in haloes below the mass resolution of the simulation. These galaxies
are called ``unresolved galaxies'' and are placed in the box in the
following way. A volume-limited sample of galaxies is generated using
the semi-analytic model, with a volume equal to that of the simulation
cube. Only galaxies which reside in haloes from the grid which are 
less massive than the resolution limit of the N-body simulation are 
considered. (Recall that the grid of halo masses used in the 
semi-analytic calculation extends to lower mass than those resolved 
in the simulation). These galaxies
are assigned to randomly selected dark matter particles which have
{\it not} been identified as members of halos identified by the 
friends-of-friends algorithm. This approach was adopted for one of the
mock catalogues used in \cite{Cole2005}. As we will see below, the
unresolved galaxies are a minority within any of the samples we
consider. They have little effect on the measured power spectrum,
producing only a modest change in the amplitude of the clustering
signal.

We can use the semi-analytic calculation carried out on the grid of
halo masses to find the completeness limit of the galaxy
catalogue in the N-body simulation. To do this, we use the galaxy formation 
calculation carried out using the grid of halo masses to compute 
the cumulative luminosity function of galaxies, starting with the brightest galaxy, 
for two cases: 1)
without any restriction on the mass of the halo which hosts the galaxy
and 2) considering only those galaxies which reside in haloes above
the resolution limit of the simulation. We then divide the second
estimate of the cumulative luminosity function by the estimate made
without any restriction on halo mass.

The completeness ratios calculated in this way are shown for $z=0, 1$ and $2$ in
Fig.~\ref{fig:comp}. The vertical lines show the magnitude limit down
to which the `resolved galaxy' catalogues are $100\%$ complete. The
lower panel shows the cumulative luminosity function in the model at
the same redshifts, with horizontal lines marking the space density of
galaxies at the sample completeness limit. (The magnitudes plotted are
observer-frame absolute magnitudes in the $R$-band. The apparent
magnitude is obtained by adding the appropriate distance modulus for
each redshift. All magnitudes are on the AB scale.) The $z=2$ sample
is complete down to $M_{R}-5\log h=-23$, or, equivalently to a space
density of $3.2\times 10^{-5}\,h^{3}\,{\rm Mpc}^{-3}$. Faintwards of this
magnitude, the completeness drops sharply to around $30-40\%$. The
situation is much more encouraging at $z=1$. Here, the galaxy
catalogue is complete to $M_{R}-5\log h=-22.3$ (corresponding to a
space density of just under $10^{-4}\,h^{3}\,{\rm Mpc}^{-3}$) and
faintwards of this there is a much more modest drop in the fraction of
galaxies resolved in the simulation. The simulation resolves around
two thirds of the space density of galaxies expected in the proposed
WFMOS survey. At $z=0$, the galaxy samples are complete to a much
higher space density, in excess of $10^{-3}\,h^{3}\,{\rm Mpc}^{-3}$.

\section{The power spectrum of galaxy clustering}

\begin{figure}
\includegraphics[width=8.5cm]{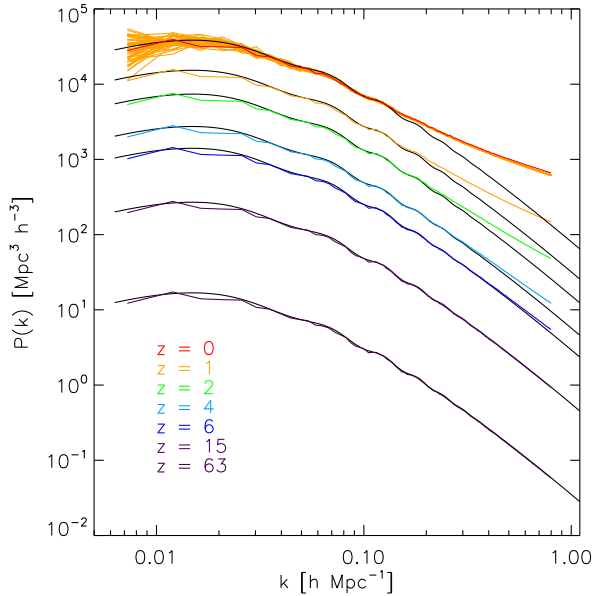} 
\caption{ The growth of
the power spectrum of density fluctuations in the dark matter, as
measured in real-space. The smooth curves show the predictions of
linear perturbation theory at the redshifts indicated by the key. The
power spectra measured in the low resolution ensemble at $z=0$ are
plotted to show the sampling variance for a simulation box of side
$1340\,h^{-1}\,$Mpc. The smallest wavenumber plotted 
corresponds to the fundamental mode in the
simulation, $2 \pi /L = 0.0469\,h^{-1}\,$Mpc. The maximum wavenumber
shown is $0.67$ times the Nyquist frequency of the FFT grid, chosen to
avoid any aliasing effects.
\label{pk:dm}}
\end{figure}

In this section we examine the various phenomena which are responsible
for changing the form of the power spectrum of galaxy clustering from
that expected in linear perturbation theory. We systematically add in
new effects and elements of sample selection, considering first the
power spectrum of the dark matter, looking at nonlinear evolution (\S
\ref{ssec:dm}) and the impact of peculiar velocities (\S
\ref{ssec:pec}), before moving onto dark matter haloes (\S
\ref{ssec:halos}) and finally to synthetic galaxy samples (\S
\ref{ssec:galaxies}). 

For completeness, we first explain some of the
terminology we use in this section. There are three types of phenomena
responsible for distorting the linear theory power spectrum: i)
non-linear growth of fluctuations, ii) redshift-space distortions and
iii) bias. Non-linear growth refers to the coupled evolution of
density fluctuations on different scales. Redshift-space distortions
describe the impact of gravitationally induced peculiar motions on the
clustering pattern. We will refer to clustering measurements as being
made in ``real-space'' or ``redshift-space''; in the latter case
peculiar motions are taken into account, as we describe in \S 3.2. The
term ``bias'' has a range of meanings in the literature. Bias is used
to describe the boost in the clustering of a particular tracer
(e.g. galaxies or clusters) relative to a reference point, which could
be the clustering of the dark matter in either linear perturbation
theory or taking into account nonlinear evolution.  One of the
earliest uses of the concept of bias was in the application of the
high peaks model to explain the enhanced clustering of Abell clusters
(\citealt{Kaiser1984}). In this model, clusters are associated with
rare peaks in the initial, Gaussian density field. The bias is defined
as the square root of the ratio of the two-point correlation function
of peaks of a certain minimum height to the clustering of the mass
expected in linear perturbation theory. When considering galaxies, it
is perhaps more natural to think in terms of a modulation of
clustering relative to that displayed by the underlying mass at the
same epoch, since galaxies populate dark matter haloes. In this case,
the galaxy clustering will be measured relative to that of the evolved
matter distribution. On large scales, these two reference points, the
clustering of the matter expected in linear perturbation theory or the
evolved clustering, should be essentially the same. We shall see later
that this is approximately the case for the scales over which we
compare clustering signals to measure bias factors.

\subsection{The nonlinear growth of matter fluctuations}
\label{ssec:dm}

\begin{figure}
  \includegraphics[width=8.5cm]{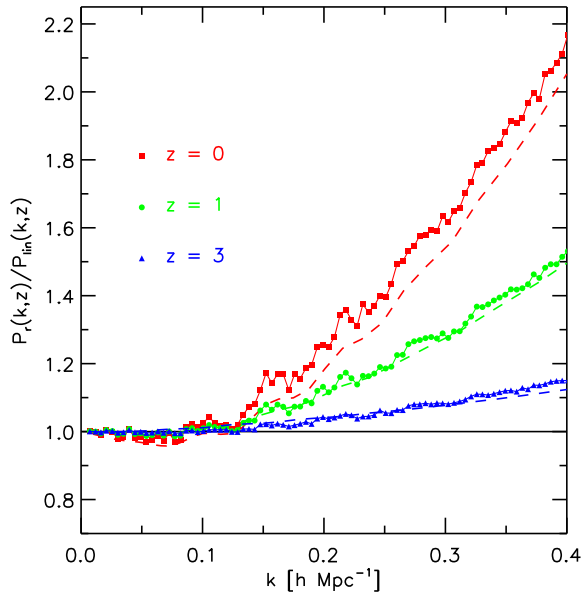} 
\caption{ The nonlinear growth of the power spectrum. Here we divide 
  the power spectrum in real-space measured at the redshift indicated by the key 
  by the power spectrum at $z=15$, after taking into account the
  change in the growth factor. Any deviation of the resulting ratio
  from unity indicates a departure from linear perturbation
  theory. The dashed lines show the same ratio as predicted using the
  ansatz of Smith et~al. (2003).
\label{fig:dm1}
}
\end{figure}

The early stages of the growth of a density fluctuation are
particularly simple to describe analytically. The fluid equations can
be written in terms of the perturbation to the density and Fourier
transformed. In the simplest case, when the density contrast $\delta
\ll 1$, the Fourier modes evolve independently of one another. This is
called linear growth. In this regime, the power spectrum changes in
amplitude with time, but not in shape. The shift in amplitude is
described by the growth factor $D$, which is a function of the
densities of matter and dark energy (as quantified by the present day
density parameters, $\Omega_{\rm M}$ and $\Omega_{\Lambda}$, for
matter and dark energy respectively) and redshift (see
\citealt{Heath1977,Peebles1980}):

\begin{equation}
P(k,z) = D^{2}(z,\Omega_{\rm M},\Omega_{\Lambda}) P(k, z=0), 
\end{equation}
where $D(z=0)=1$. 

We plot the power spectrum of the dark matter in real-space measured
from the {\tt BASICC} at different output redshifts in Fig.~\ref{pk:dm}. The
approximately linear growth of the power spectrum is readily apparent
on large scales (low k). In an Einstein - de Sitter universe
($\Omega_{\rm M}=1$), the growth factor is equal to the expansion
factor. If dark energy plays a role in setting the rate at which the
universe expands, the growth of fluctuations is suppressed relative to
the Einstein - de Sitter case at late times. The {\tt BASICC} started at
$z_{\rm s}=63$, so if $\Omega_{\rm M}=1$, we would expect to see the
power spectrum grow in amplitude by a factor of $(1+z_{\rm
s})^{2}=4096$ by $z=0$. Using the approximate formula provided by
\cite{Carroll1992}, we expect a suppression in the growth of the power
by a factor of 0.5537 for the cosmological parameters used in the
simulation. This gives an overall growth in power from the initial
conditions to the present of a factor of 2268. This agrees to within
0.6\% with the factor expected from a direct numerical integration of
the equation giving the growth factor (eqns. 28 and 9 from
\citealt{Carroll1992}), which gives 2281.01. In the simulation, we
find that the power in the fundamental mode grows by a factor of
2285.21 from the initial conditions at $z=63$ to $z=0$, which agrees 
with the growth predicted by linear perturbation theory to $0.02\%$. 

Fig.~\ref{pk:dm} shows that the growth of the power spectrum is clearly not linear at high 
wavenumbers. The shape of the spectrum at high $k$ at late times is different from that 
at high redshift, because the growth of modes of different $k$ becomes coupled. This 
behaviour can be followed to some extent using second- and higher-order perturbation theory 
(\citealt{Peebles1980, Baugh1994, Jain1994, Crocce2006a}). However, as the 
density contrast approaches unity, second-order perturbation theory breaks down 
(\citealt{Baugh1994}). The coupled evolution of the Fourier modes starts on surprisingly 
large scales, which demonstrates the necessity of a large volume simulation to accurately 
follow the development of the power spectrum (Smith, Scoccimarro \& Sheth 2006). 
This can be seen more clearly if we divide the 
measured spectrum by the growth expected according to linear perturbation theory, as is done 
approximately in Fig.~\ref{fig:dm1}. In this plot, we have divided the power spectra measured 
from the simulation by the spectrum measured at $z=15$, scaled by the square of the appropriate growth 
factor. This reduces the noise in the ratio arising from the finite number of modes realized 
at small wavenumbers in the simulation volume (\citealt{Baugh1994, Millennium}). 
Any deviation away from unity signifies a departure from linear perturbation theory due to 
coupling between modes.  The ratio shows a characteristic dip at low $k$, i.e. less power than 
expected in linear theory, before showing a strong enhancement at higher wavenumbers 
(\citealt{Baugh1994}). It is remarkable that the transition between a deficit and excess 
of power happens at the same wavenumber, $k \sim 0.1 h {\rm Mpc}^{-1}$, at different epochs. 
The suppression in power at low $k$, on the order of a 3\%, is not as strong as that seen in 
an Einstein - de Sitter universe (see figure 4 of \citealt{Baugh1994}). Nevertheless, this drives 
the spectacular boost in power seen at higher wavenumbers. The dip in power is largest around 
$k \sim 0.05 h {\rm Mpc}^{-1}$, which corresponds to a length scale of $2 \pi /k \sim 125\,h^{-1}\,{\rm Mpc}$, 
close to the wavelength of the acoustic oscillations. Several authors have proposed ansatzes which 
transform the linear perturbation theory power spectrum into the non-linear power 
spectrum (e.g. \citealt{Hamilton1991, 
PD1994, PD1996, Smith2003}). We plot the predictions of the model proposed 
by \cite{Smith2003} in Fig.~\ref{fig:dm1} using dashed lines. The ratio is computed by dividing the power 
spectrum at the epoch of interest by the suitably scaled prediction of the model for z=15. The 
agreement is excellent at high redshift. At $z=0$, at higher wavenumbers, the \cite{Smith2003} 
formula recovers the simulation results to within 5\% over the range plotted. 

\subsection{The impact of redshift-space distortions on the power spectrum}
\label{ssec:pec}

\begin{figure}
  \includegraphics[width=8.5cm]{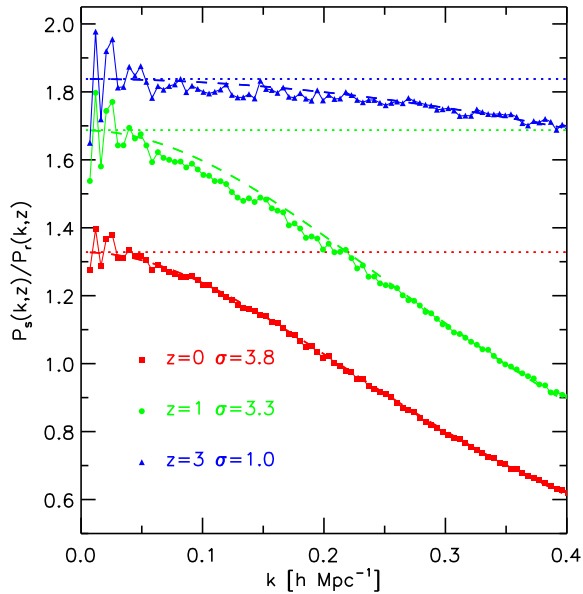}
   \caption{
The ratio of the power spectrum measured for the dark matter in redshift-space, i.e. 
including the impact of peculiar motions in the distance determination, to the power 
spectrum measured in real-space. The deviation from unity shows the redshift-space 
distortion to the nonlinear power spectrum. The results are shown for selected output 
redshifts, as indicated by the key. The horizontal dotted lines indicate the boost in 
the redshift-space power expected due to coherent flows, as predicted by Eq.~\ref{boost}.
The dashed lines show a simple fit to the distortions (see eq \ref{damp}). 
\label{fig:dmzs}
}
\end{figure}

In a spectroscopic galaxy survey, the radial distance to an object is inferred from its measured redshift. The 
shift in the spectral features of the galaxy is produced by two contributions to its the apparent velocity: 
the expansion of the universe, which is responsible for the Hubble flow at the true distance to the galaxy, 
and local inhomogeneities in the gravitational field around the object, which generate an additional, 
``peculiar'' velocity. Since we cannot correct a priori for the effects of the local gravitational field 
when inferring the radial distance from the Hubble law and the measured redshift, an error is made in 
the distance determination. The impact of such errors on the form of the measured power spectrum of 
clustering is called the redshift-space distortion.  

Peculiar motions display two extremes which produce different types of distortion to the power spectrum: 
i) On large scales, coherent bulk flows out of voids and into overdense regions lead to an enhancement in 
the density inferred in redshift-space, and hence to a boost in the recovered power. \cite{Kaiser1987} derived 
a formula for the enhancement of the spherically averaged power, under the assumption of linear perturbation theory for an observer 
situated at infinity (the plane parallel approximation): 
\begin{equation}\label{boost}
f = \frac{P_{\rm s}(k)}{P_{\rm r}(k)} = (1 + \frac{2}{3}\beta + \frac{1}{5}\beta^2 ) ,
\end{equation}
where $P_{\rm s}(k)$ is the power spectrum in redshift-space, $P_{\rm r}(k)$ is the spectrum in real-space 
and $\beta = \left({\rm d} \log \delta / {\rm d} \log a \right) /b \simeq \Omega_{\rm M}^{0.6}(z)/b$, where b is the bias factor ($b=1$ for the dark matter; for a discussion of the dependence of the growth 
factor on $\Omega_{\rm M}$, see Linder 2005; Linder \& Cahn 2007).
ii) On small scales, the random motions of objects inside virialized dark matter haloes cause structures to 
appear elongated when viewed in redshift-space, leading to a damping of the power. 
\cite{PD1994} discussed a model for the redshift-space power spectrum, 
which takes into account both limits of peculiar motions (see also \citealt{Scoccimarro2004}).

Fig.~\ref{fig:dmzs} shows the ratio of the power spectrum measured for the dark matter in 
redshift-space to that measured in real-space, at redshifts $z=3,1$ and 0. The dotted lines indicate the boost 
expected in the redshift-space power, computed using the expression in Eq.~\ref{boost} 
(\citealt{Kaiser1987}). 
This factor changes with redshift because the matter density parameter is changing. Fig.~\ref{fig:dmzs} 
shows that this behaviour is only approached asymptotically, on scales in excess of $100\,h^{-1}\,$Mpc. 
At higher wavenumbers, the power measured in redshift-space is suppressed by random motions. 
The dashed lines in this plot show a simple fit to this ratio 
\begin{equation}\label{damp}
f = \frac{P_{\rm s}(k)}{P_{\rm r}(k)} = (1 + \frac{2}{3}\beta + \frac{1}{5}\beta^2 ) (1 + k^2 \sigma^2)^{-1}, 
\end{equation}
where $\sigma$ is a free parameter, which is loosely connected to the pairwise velocity dispersion. 
The degree of damping grows between $z=3$ and $z=1$, but changes relatively little by $z=0$. 
We shall see in later sections that the form of the redshift-space distortion to the power spectrum 
depends on the type of object under consideration. 

\subsection{The power spectrum of dark matter halos in real and redshift-space}
\label{ssec:halos}

\begin{figure*}
\includegraphics[width=\textwidth]{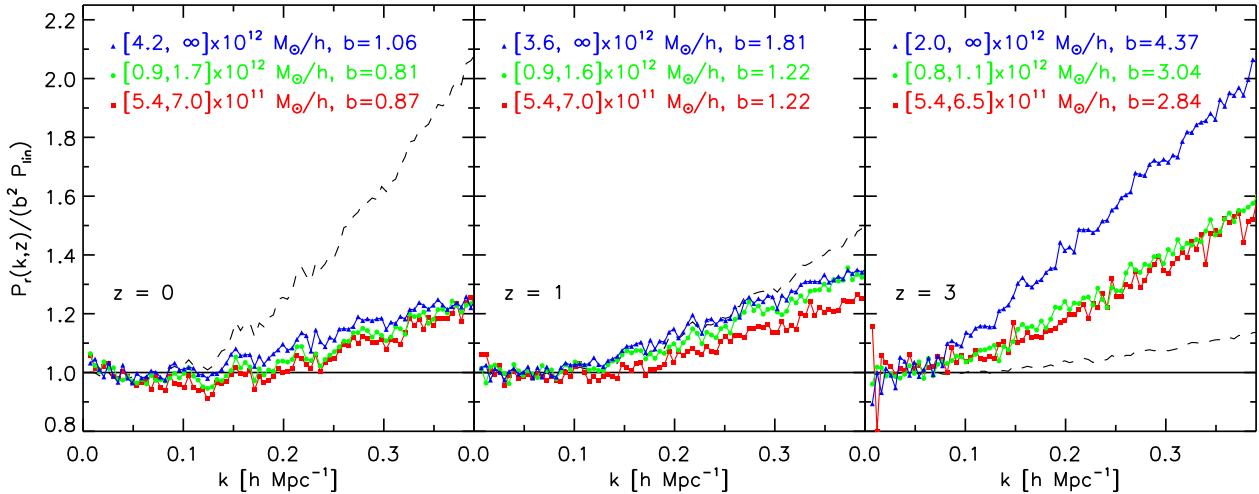}
\caption{
The power spectrum of dark matter haloes measured in real-space compared to a scaled version of the 
prediction of linear perturbation theory, which takes into account the growth factor and an effective bias 
computed on large scales $k < 0.1 h {\rm Mpc}^{-1}$. Each panel corresponds to a different output redshift. 
Different mass samples are considered, as indicated by the key, which correspond to low, average and high 
masses, defined in terms of the average halo mass present at each output time. The black dashed line shows 
the real-space power spectrum of the mass divided by the appropriate linear perturbation theory prediction. 
\label{fig:halosr}}
\end{figure*}

\begin{figure*}
\includegraphics[width=\textwidth]{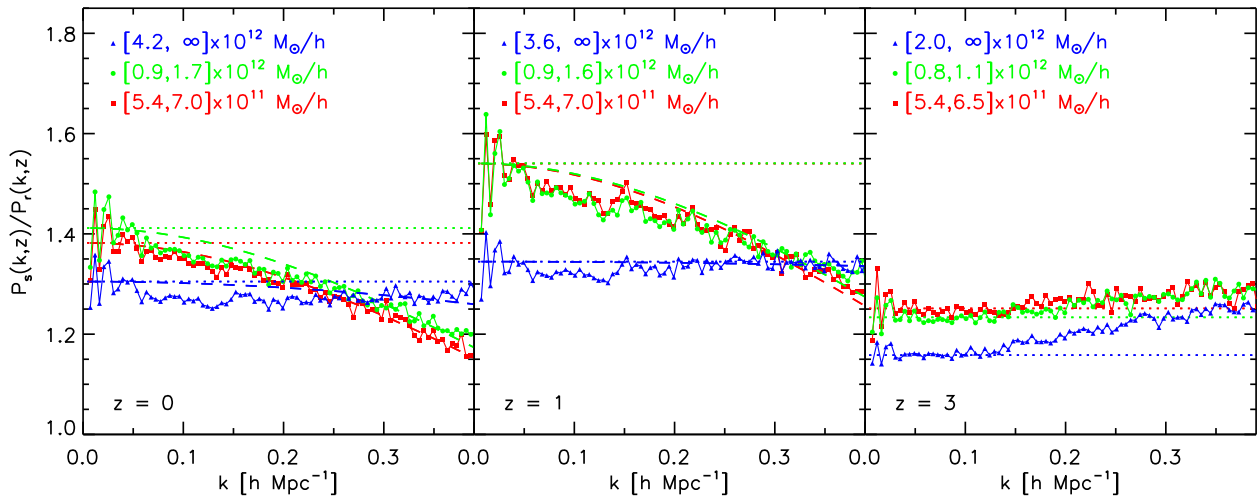}
\caption{
The power spectrum of dark matter haloes measured in redshift-space divided by the power spectrum measured 
in real-space for the same sample. Each panel corresponds to a different output redshift. 
Different mass samples are considered, as indicated by the key, which correspond to low, average and high 
masses, defined in terms of the average halo mass present at each output time. 
The horizontal dotted lines show the expected ratio for the boost in the amplitude of the redshift-space 
power spectrum due to coherent flows, computed using an effective bias factor estimated on large scales.  
The dashed lines show the best fit model of Eq.~\ref{damp}, which turns out to be a poor description of 
the redshift-space distortions. No suitable fits were obtained at $z=3$. 
\label{fig:halosz}}
\end{figure*}

In modern theories of galaxy formation, dark matter haloes play host to galaxies. 
It is therefore instructive to compare the power spectra measured for 
different samples of haloes to that of the dark matter as a step towards 
understanding the power spectrum of galaxies. 

A common conception is that the clustering of haloes is a scaled version of the clustering 
of the underlying mass, with the shift in clustering amplitude quantified in terms of a 
bias factor, $b$, where $b^2  = P_{\rm{halos}}/P_{\rm{dm}}$ (\citealt{Cole1989, MoWhite1996}). 
As we commented earlier, since we use the dark matter power spectrum on large scales to define 
a bias, this is approximately the same as using the linear perturbation theory spectrum. 
Many authors have tested analytical prescriptions for computing 
the bias parameter using extensions of the theory of \cite{PressSchechter1974} 
(e.g. Mo, Jing \& White 1997; Sheth, Mo \& Tormen 2000; \citealt{Jing1998, Governato1999, Colberg2000, SeljakWarren2004}).
In the extended Press-Schechter theory, the bias is only a function of halo mass and redshift. 
However, recent analyses of high resolution, large volume simulations have revealed 
some dependence of halo clustering on a second parameter besides mass, such as the halo's 
formation redshift or concentration parameter (\citealt{Gao2005, Harker2006, Wechsler2006}). 

In Fig.~\ref{fig:halosr}, we show that this simple picture, in which the clustering 
of haloes is a shifted version of that of the dark matter, is actually a poor approximation 
to what we find in the simulation. We show the ratio of the power spectrum of a sample 
of dark matter haloes measured in real-space to a scaled version of the linear perturbation 
theory power spectrum. The amplitude of the linear theory spectrum used in 
the ratio takes into account the growth factor appropriate to the output redshift 
and an effective bias, which is set by matching the linear theory prediction for the mass spectrum 
to the measured halo spectrum on large scales, i.e. for wavenumbers in the range 
$0.0046 < (k/h {\rm Mpc}^{-1}) < 0.1$. Each panel in Fig.~\ref{fig:halosr} corresponds 
to a different output redshift from the simulation. For each redshift, we have defined three 
samples of dark matter haloes, which contain the same number of objects. The mass intervals 
are set relative to the average halo mass present in the respective outputs, with 
``low'', ``mean'' and ``high'' mass  samples considered. Each of these contains 
20\% of the total number of haloes present at each epoch, with the  
mass ranges used at each redshift indicated on the keys. The effective bias factors of the halo 
samples are also written in the key. For comparison, the dashed line in each panel shows 
the corresponding ratio for the dark matter. 

Fig.~\ref{fig:halosr} shows that at $z=3$, all of the haloes considered have effective 
biases much greater 
than unity, indicating they are more strongly clustered than the mass. 
This situation is reversed at $z=0$. At this epoch, the halo mass resolution 
of the {\tt BASICC} is smaller than the corresponding 
value of $M_{*}$\footnote{$M_{*}$ is a characteristic mass scale 
defined as the mass within a sphere for which the {\it rms} variance 
in linear perturbation theory is $\sigma (M) = \delta_{\rm crit}(z)$, 
where $\delta_{\rm crit}$ is the extrapolated critical linear overdensity 
given by the spherical collapse model at redshift $z$.} ($=5.78 \times 10^{12}\,h^{-1} \, M_{\odot}$ at z=0). 
The $z=0$ samples have a bias of unity or smaller. In addition to the difference 
in the effective bias parameters, the shape of the spectrum of the haloes in these extremes is 
also different (see also Smith et~al. 2006). The plot shows the shape of the power spectrum, 
after accounting for the effective 
bias on large scales. Any difference between the curves plotted for the haloes and that for the dark 
matter (dashed line) shows a difference in the clustering signal over and above that quantified by a constant 
effective bias. 
Similar behaviour was found for samples of cluster mass haloes 
in the Hubble Volume simulation by \cite{Angulo2005}. 

We now consider the clustering of haloes as viewed in
redshift-space, taking the centre of mass velocity of 
the halo as its peculiar velocity. In Fig.~\ref{fig:halosz}, we plot the ratio of the
redshift-space power spectrum for the halo samples used in
Fig.~\ref{fig:halosr} to the power spectrum measured in real-space. As
we did before for the case of the dark matter (Fig.~\ref{fig:dmzs}),
we indicate the boost in power expected on large scales (small $k$)
due to coherent bulk flows of haloes. The boost is calculated from
Eq.~\ref{boost} using the effective bias of the halo sample. The plot
shows that the redshift-space power spectrum at low wavenumbers is in
reasonable agreement with this simple model. However, a range of
behaviour is seen at higher wavenumbers. For haloes comparable to
$M_{*}$, the boost in power in redshift-space is less than predicted
by Eq.~\ref{boost}. For the more extreme, massive haloes, there is
actually more power in redshift-space than is suggested by Kaiser's
formula. This ``excess'' power was previously noted by
\cite{Padilla2002} and \cite{Angulo2005}. The Kaiser formula assumes
linear perturbation theory and breaks down in the case of objects with
strongly nonlinear clustering. In the case of the less extreme haloes,
the reduction in power is {\it not} due to virialized motions of
haloes within larger structures. The halo finder we have used is
designed to return an overdensity corresponding to virialized
structures and not substructures. If the haloes were really part of a
larger structure and were executing random motions, the group finder
would simply have lumped them together as one larger structure. We are
perhaps seeing instead haloes that have started to merge with one
another, and whose motions have broken away from a coherent large
scale flow. We know of no analytical description of the redshift-space
clustering of dark matter haloes which explains this behaviour.

\subsection{The power spectrum of galaxies}
\label{ssec:galaxies}

\begin{figure*}
\includegraphics[width=14.5cm]{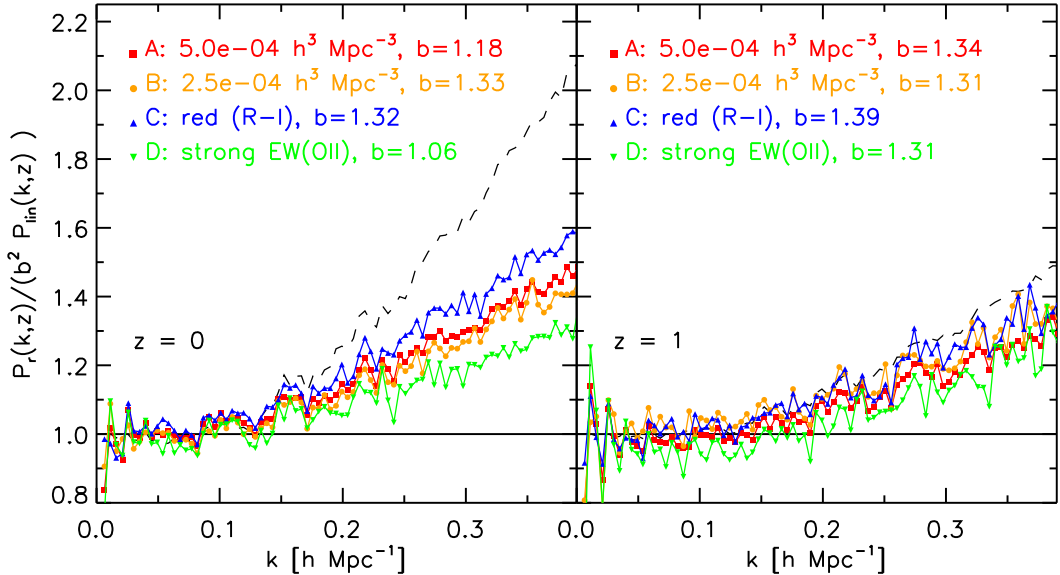}
\caption{ 
The power spectrum of different galaxy samples measured in real-space, divided by 
the square of an effective bias parameter and the appropriately scaled linear 
perturbation theory power spectrum. The sample definition and the value of the 
effective bias used are given by the key. The power spectrum of the dark matter 
spectrum in real-space, also divided by the linear perturbation theory spectrum, 
is shown by the black dashed line. The left hand panel shows the ratios at $z=0$ and 
the right hand panel at $z=1$. 
\label{fig:galr}}
\end{figure*}

\begin{figure*}
\includegraphics[width=14.5cm]{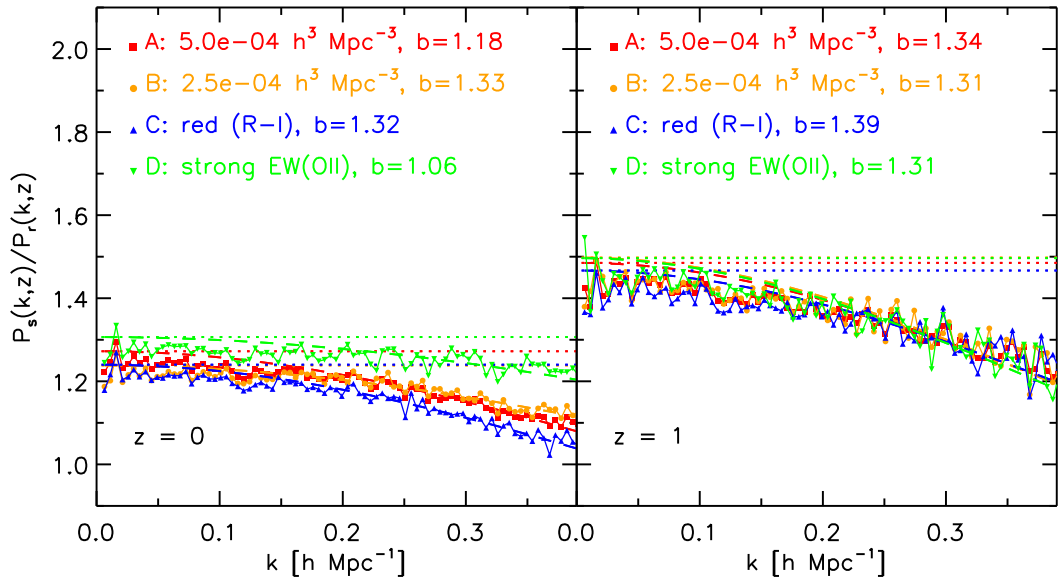}
\caption{ 
The ratio of the power spectrum of galaxies measured in redshift-space to that 
in real-space, at $z=0$ (left) and $z=1$ (right). The samples are defined by 
the key in each panel. The dotted horizontal lines show the 
predictions of Eq.~\ref{boost} for the various samples. 
\label{fig:galz}}
\end{figure*}

The galaxy power spectrum can be very different from the power spectrum of a sample 
of dark matter haloes. The way in which the galaxies are distributed among haloes 
changes the form of the power spectrum. In a mass-limited sample of haloes, the contribution 
of each halo to the power spectrum can be determined through its space density, which 
acts as a weighting factor when computing the contribution of the halo to the 
clustering signal. The number of galaxies per halo acts to modify this weight e.g. 
more massive haloes could contain more galaxies than less massive haloes. Furthermore, 
the presence of satellite galaxies within a halo means that one expects to see a damping 
in power on small scales in redshift-space, due to the random motions of the satellites 
within the virialized dark halo. The precise modification of the power spectrum depends in 
detail on how galaxies populate dark matter haloes. 
As we discussed in \S 2.2, we have carried out an {\it ab initio} calculation of the number 
of galaxies per halo, using a semi-analytic model of galaxy formation.
We are able to predict observable properties of galaxies, such as broadband 
magnitudes and the strength of emission lines. 
We consider a range of galaxy samples, defined either by a magnitude 
limit alone (set in the R-band) or by combining an R-band magnitude limit 
with a colour selection (in R-I) or a cut 
on the strength of the OII[3727] emission line: 
\begin{itemize}
\item  Sample A: magnitude-limited to reach a space density of 
$5 \times 10^{-4}\,h^{3}\,{\rm Mpc}^{-3}$. 

\item  Sample B: magnitude-limited to reach half the space density of 
sample A, i.e. $2.5 \times 10^{-4}\,h^{3}\,{\rm Mpc}^{-3}$. 

\item  Sample C. The reddest 50\% of galaxies from sample A, 
using the $R-I$ colour.

\item  Sample D. The 50\% of galaxies from sample A with the 
strongest emission lines, using the equivalent width of OII[3727].

\item  Sample E. The bluest 50\% of galaxies from sample A, 
using the $R-I$ colour. 

\item  Sample F. The 50\% of galaxies from sample A with the 
weakest emission lines, using the equivalent width of OII[3727].

\end{itemize}

The power spectra measured in real-space from the various galaxy samples are plotted in 
Fig.~\ref{fig:galr}. 
The spectra have been divided by the linear perturbation theory power spectrum 
multiplied by the square of an effective 
bias factor, which was estimated by comparing the galaxy spectra to the power spectrum 
measured for the dark matter for wavenumbers $k < 0.1 h {\rm Mpc}^{-1}$. In all cases, 
for the space densities we have chosen, the effective bias factors estimated for the 
samples are modest. For comparison, the ratio of the power spectrum of the dark matter 
in real-space to the linear theory prediction is also plotted, using a dashed line. 
The deviation of the dashed line from unity shows where nonlinear effects are important for 
the dark matter. Any differences between the plotted ratios for galaxies and 
mass indicate a scale dependent bias. The comparison between the dashed and 
solid curves in Fig~\ref{fig:galr} shows that a constant bias is only a good 
approximation on large scales, $k < 0.15 h {\rm Mpc}^{-1}$.

The redshift-space distortion in the galaxy power spectrum is shown in Fig~\ref{fig:galz}, 
where we plot the ratio of the redshift-space spectrum to the real-space spectrum for 
the galaxy samples shown in Fig.~\ref{fig:galr}. The horizontal lines show the Kaiser 
boost (Eq.~\ref{boost}) expected for the effective bias of the galaxy sample. This ratio 
is only attained on the very largest scales and seems to be an overestimate of the size 
of the effect at $z=1$. The damping of the power on intermediate and small scales is readily 
apparent and, unlike the case with dark matter haloes, is well described by the form given 
in Eq.~\ref{damp}. 

\section{Constraining the Dark Energy Equation of state}

\begin{figure}
  \includegraphics[width=8.5cm]{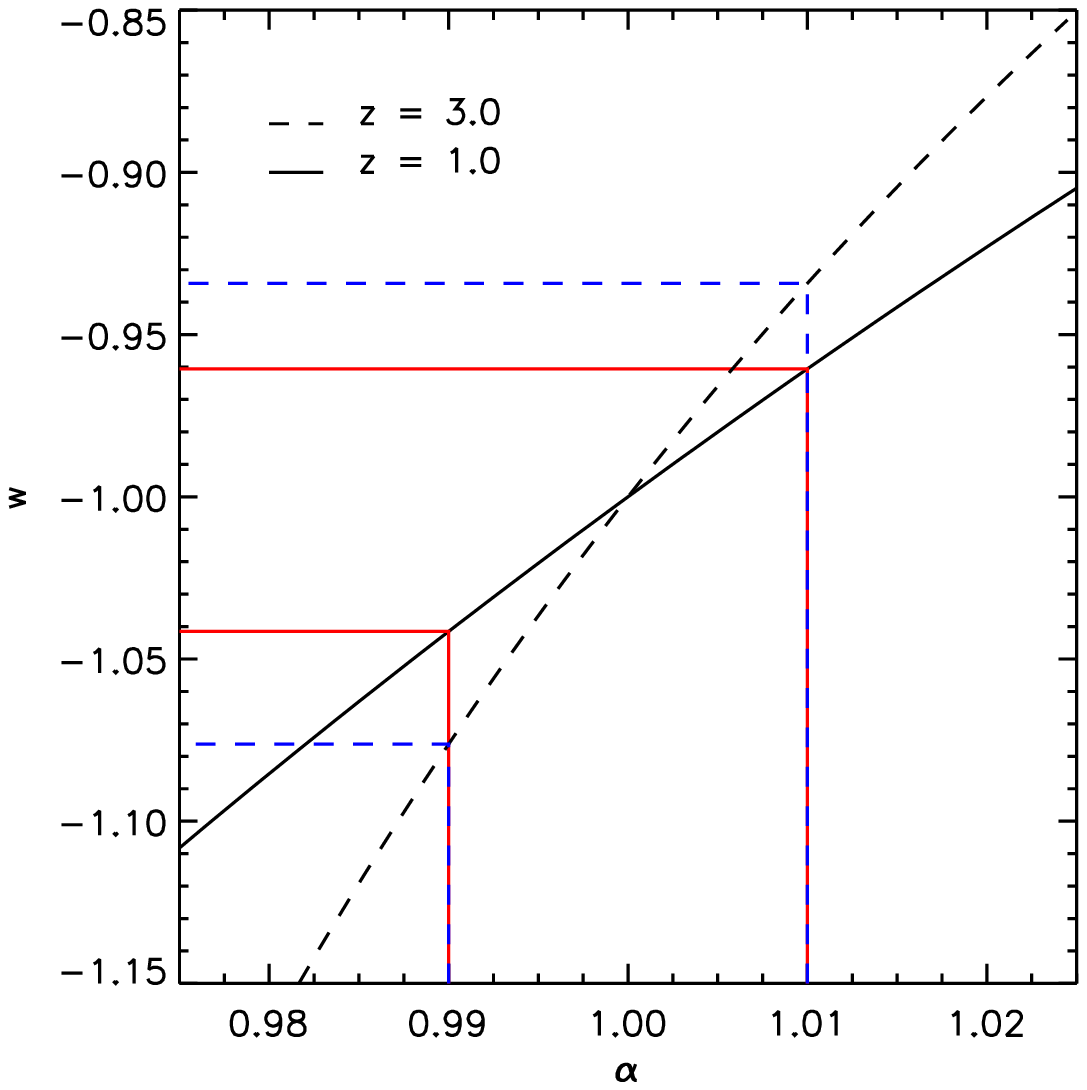}
  \includegraphics[width=8.5cm]{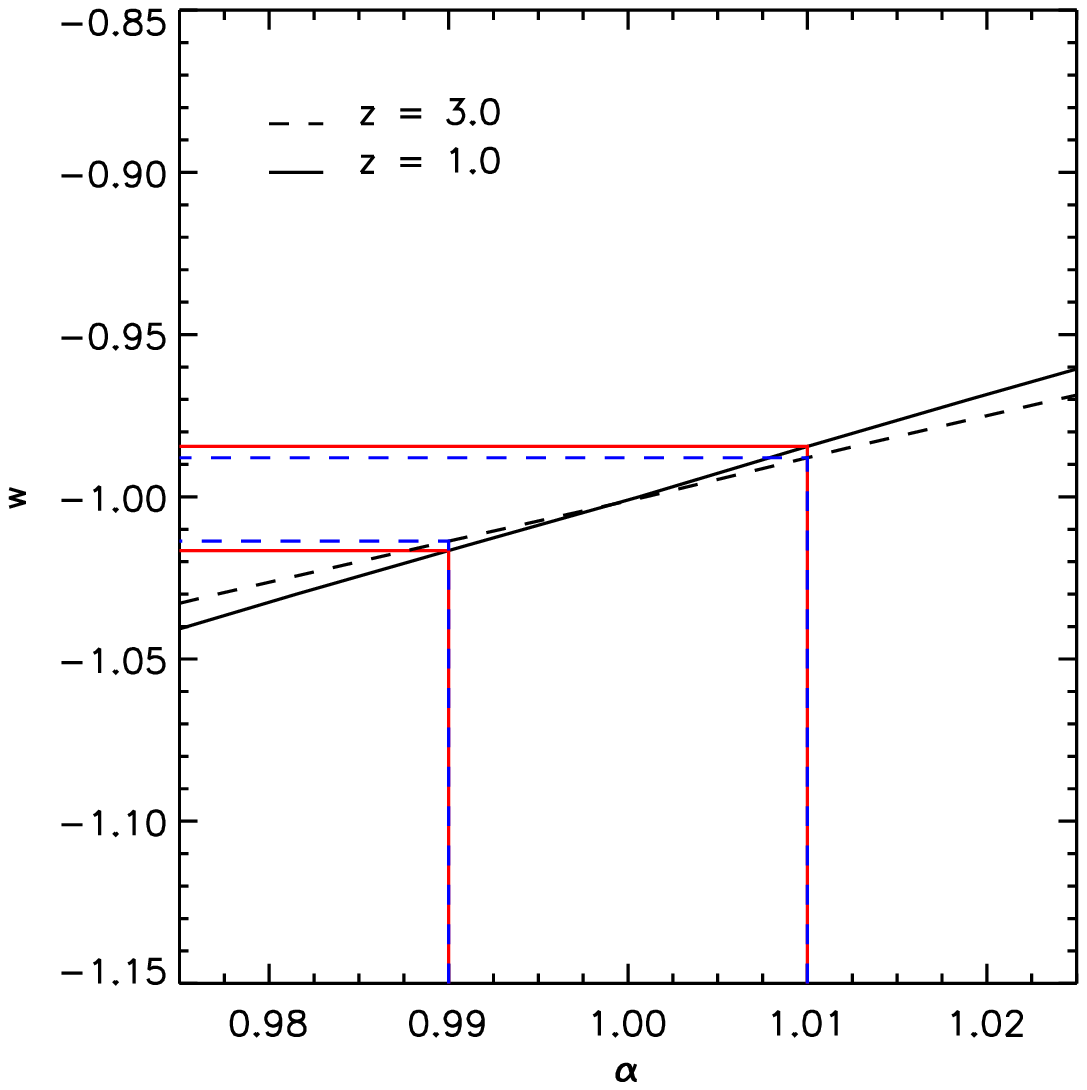}
   \caption{
The relation between the dark energy equation of state parameter, $w$, and 
the scale factor, $\alpha$, defined by Eq.~\ref{eq:alpha}, for perturbations 
in the equation of state around $w_{\rm true}=-1$. Two cases are shown. In the 
upper panel, the values of the other cosmological parameters are kept fixed. 
In the lower panel, the ratio of the sound horizon scale to the angular 
diameter distance to the last scattering surface is held fixed. 
The relation between $\alpha$ and $w$ is shown for $z=1$ (solid lines) 
and $z=3$ (dashed lines). The horizontal and vertical lines guide the eye 
to show how a $1\%$ error in $\alpha$ translates into an error in $w$. 
\label{fig:walpha}
}
\end{figure}

In this section we outline the procedures we follow to place constraints 
on the dark energy equation of state parameter, $w$, by measuring the 
length scale imprinted by baryonic acoustic oscillations on the power 
spectrum of the various tracers of the density field. The transformation 
of a measurement of a distance scale into a constraint on $w$ requires 
various approximations to be made, and depends upon the survey in question and 
upon the time variation assumed for the dark energy. Nevertheless it is 
instructive to go through this exercise, bearing these caveats in mind, to 
get a feel for how well future experiments will be able to measure $w$ for 
the case of a constant equation of state.

The form of the power spectrum of density fluctuations contains information 
about basic cosmological parameters, and measurements of the galaxy power 
spectrum on large scales have been exploited to extract the values of these 
parameters (e.g. \citealt{Cole2005, Sanchez2006, Tegmark2006, Padmanabhan2007, 
Percival2007}).  
The apparent scale of features in the power spectrum offers another route to 
constrain selected cosmological parameters through the dependence of the 
distances parallel and perpendicular to the line of sight on the matter 
density parameter, $\Omega_{\rm M}$, the dark energy density parameter, 
$\Omega_{\rm DE}$, the dark energy equation of state parameter, $w$ and the 
Hubble constant. For such an approach to work, we either need to know the true 
physical scale of a particular feature in the power spectrum beforehand or to 
compare the relative size of a feature when measured parallel and perpendicular 
to the line of sight (Alcock \& Paczynski 1978). 
The baryonic oscillations present a promising candidate for such a feature. 
If we assume for the sake of argument that the cosmological parameters, 
apart from the equation of state of the dark energy, are well constrained, 
then the scale of the acoustic oscillations becomes a standard ruler. 
These features are expected on smaller scales than the turnover and have 
already been seen in current surveys at low redshift, although at too 
low a signal-to-noise ratio to use in isolation to extract a competitive 
constraint on the dark energy equation of state (Cole et~al. 2005; Eisenstein et~al. 2005).  

We can see how the value of the equation of state parameter parameter of the 
dark energy influences the form of the BAO with the following simple 
argument. To measure the power spectrum of galaxy clustering, we need to convert 
the angular positions and redshifts of the galaxies into comoving spatial separations. 
This requires a choice to be made for values of the cosmological parameters, 
including $w$. In our case, we set the parameters equal to the values used in the 
N-body simulations, with $w=w_{\rm true} = -1$ for the particular case we have run. 
The effect of a change in the value of $w$, $w_{\rm assumed} = 
w_{\rm true} + \delta w$ is to change the separations between pairs of 
galaxies, which leads to a change in the appearance of the power spectrum.  
For small perturbations away from the true equation of state, we assume that 
the alteration in the measured power spectrum can be represented by a rescaling 
of the wavenumber from $k_{\rm true}$ to $k_{\rm app}$. The ratio of these wavenumbers 
gives a ``stretch'' parameter, $\alpha$, which describes the change in the recovered 
oscillation scale: 
\begin{equation}
\alpha = \frac{k_{\rm app}}{k_{\rm true}}. 
\end{equation} 
If $w_{\rm assumed} = w_{\rm true}$, then there is no shift in the BAO in the 
estimated power spectrum and $\alpha=1$. 
In the case of a wide-angle, deep galaxy survey with spectroscopic redshifts, 
the stretch parameter can be approximated by: 
\begin{equation}
\alpha \approx \left( \frac{ D_A(z,w_{\rm assumed}) }{ D_A(z,w_{\rm true}) } \right)^{-2/3} 
\left( \frac{ H(z,w_{\rm true}) }{ H(z,w_{\rm assumed}) } \right)^{1/3},
\label{eq:alpha}
\end{equation}
where 
\begin{eqnarray}
H(z,w) &=& H_0 \left[ \Omega_m (1+z)^3+\Omega_{\rm DE} (1+z)^{3(1+w)} \right]^{1/2} \\
D_A(z,w) &=& \frac{c}{1+z}\int_{0}^{z}\frac{\rm{dz}}{H(z)}.
\label{eq:hz}
\end{eqnarray} 
The values of the exponents in Eq.~\ref{eq:alpha}, $2/3$ for the distance transverse 
to the line of sight and $1/3$ for the distance parallel to the line of sight 
are motivated by the number of cartesian components in these directions 
(e.g. \citealt{Eisenstein2005}). The precise value of these exponents will 
depend upon the geometry and construction of the galaxy survey. For example, 
in a survey which relies upon photometric redshifts, the exponent parallel 
to the line of sight would be greatly reduced and it would be beneficial to 
compute the power spectrum transverse to the line of sight.  
Note that in Eqs.~9 and 10 we assume that $w$ is independent of redshift. 
There are many models in which $w$ is a function of redshift. In this case, 
the exponent of $\Omega_{\rm DE}$ in the expression for the Hubble parameter 
(Eq.~9) would be replaced by an integral over $w(z)$. 

It is instructive to see how the constraints on $\alpha$ translate 
into limits on the value of $w$. We can do this approximately using 
Eq.~\ref{eq:alpha}, for the case of a redshift independent 
equation of state, considering perturbations around $w_{\rm true}=-1$. 
We consider two illustrative cases: a ``pessimistic'' case in which we consider 
the constraints from BAO in isolation from any other data which constrains 
the cosmological parameters and an ``optimistic'' case, in which we 
perturb $w$ and only consider cosmological models that give similar 
predictions for the CMB.\footnote{We acknowledge the referee for 
suggesting this second case to us and for encouraging us to perform 
the calculation.} The translation in the pessimistic case is shown 
in the upper panel of Fig.~\ref{fig:walpha} for two different redshifts. 
Here we have assumed fixed values for $\Omega_{\rm M}$ and $\Omega_\Lambda$ 
and we have not marginalized over these parameters. This is the case 
discussed most commonly in the literature. Under these conditions, 
at $z=1$, a $1\%$ error in $\alpha$ corresponds approximately to 
a $4\%$ error in the value of $w$. At $z=3$, the boost is about 50\% larger, 
with $\delta w \approx 6 \delta \alpha$. 

In the ``optimistic'' case, we only consider models which give 
the same angular location for the first peak in the CMB spectrum. 
Hence, when the value of $w$ is perturbed, we restrict our attention to 
those models which give the same ratio of the sound horizon scale 
to the angular diameter distance to the last scattering surface 
as our default cosmology. Given the parametric forms quoted for 
these distances by Eisenstein \& Hu (1998), this is equivalent 
to keeping $\Omega_{\rm b}/\Omega_{\rm M}$ and $h$ fixed, and varying 
$\Omega_{\rm M}$. We have 
called this case ``optimistic'' because it does not include any 
error on the fixed parameters. In this scenario, shown 
in the lower panel of Fig.~\ref{fig:walpha}, the error on $w$ is 
now only around 50\% larger than the corresponding error on $\alpha$.

We now explore two of the approaches which have been advocated in the literature 
to measure the value of $w$. Both methods involve making fits to the ratio of a 
measured power spectrum divided by a smooth reference spectrum. In the first approach, 
a parametric form is assumed for the ratio (\citealt{BG2003}). 
The second approach is more general as it does not assume a specific form for the 
ratio, but instead uses the linear perturbation theory power spectrum without any 
further approximations (\citealt{Percival2007}; see also \citealt{Eisenstein2005}). 
We shall henceforth refer to these methods as the parametric and general schemes 
respectively. In their original forms, there are also differences in the way in which 
a ``featureless'' reference spectrum is constructed, as we will briefly discuss when 
describing these approaches below.  

Blake \& Glazebrook (2003; see also Glazebrook \& Blake 2005) studied the feasibility of extracting measurements of 
the acoustic oscillations from forthcoming galaxy surveys using linear perturbation theory. 
Their starting point is to divide the power spectrum, including the imprint 
of baryons, divided by a smooth reference spectrum which is chosen to be free from any 
signature of acoustic oscillations. This method therefore does not use any of the information 
contained in the overall shape of the power spectrum, which Blake \& Glazebrook argue could 
be susceptible to large scale gradients arising from the effects we discussed in Section 3, 
such as galaxy bias or redshift-space distortions. Instead, they focused on the location 
and amplitude of the acoustic oscillations. The smooth reference spectrum is obtained using the 
zero-baryon transfer function written down by \cite{EH1998}. The 
parametric form suggested by Blake \& Glazebrook as a fit to the resulting ratio is a Taylor expansion 
of the ratio of a power spectrum for cold dark matter plus a small baryonic component, 
divided by a pure cold dark matter power spectrum. The sound horizon, which is a free parameter 
in their method, is treated as the oscillation wavelength in this parametric form. This is an  
approximation, as the wavelength of the acoustic oscillations actually changes with 
wavenumber, albeit slowly, and is therefore not a constant (see eqn.~22 of \citealt{EH1998}). 
Some authors have criticized this approach due to the sensitivity of the ratio to the choice of 
the reference power spectrum. \cite{Angulo2005} describe how realistic power spectra, which 
include nonlinear growth, bias effects and redshift-space distortions, require a ``linearization'' 
process before they become adequately described by the parametric form put forward 
by Blake \& Glazebrook. 
Due to the sensitivity of the ratio to the choice of reference spectrum at low wavenumbers, 
\cite{Koehler2006}) proposed ignoring power spectrum measurements below 
$k \sim 0.05 h{\rm Mpc}^{-1}$ to avoid this problem (although we note that they also discuss 
a different approach to measuring the equation of state parameter). 

\cite{Percival2007} proposed a new technique which has a number of appealing features compared 
with that of Blake \& Glazebrook. Firstly, the shortcut of fitting an 
approximate parametric form to the ratio of the measured power spectrum to a reference is dropped 
in favour of using a full linear perturbation theory power spectrum (with a modification; see later) 
to model the ratio. This is completely general, and permits one to use the most accurate description 
available of the linear perturbation theory power spectrum, such as the tabulated output of {\tt CAMB}. 
Secondly, the reference power spectrum is defined separately in the case of the data and the 
linear theory model, by using a coarse rebinning of the relevant power spectrum. 
The reference is constructed using a spline fit to a reduced number of wavenumber 
bins over the range in which the spectrum in question is defined. 
Thus, any deviations in the general form of the measured spectrum away from linear 
theory are naturally accounted for in the reference spectrum. Thirdly, Percival et~al. 
allow for a damping of the amplitude of the oscillations in the theoretical ratio 
beyond some wavenumber, which is treated as a free parameter in their fit. 
The quality of the fits is dramatically improved when damping of the higher harmonics is allowed. 
Percival et~al. applied their method to extract the matter density parameter from the power spectrum 
of luminous red galaxies in the SDSS. 

The majority of the results we present are obtained using the general method suggested by Percival et~al. 
For completeness, and because Percival et~al. did not actually apply their method to the extraction 
of the equation of state parameter, we set out the general approach step-by-step below: 

\begin{enumerate}

\item[1.] 
A smooth reference spectrum (i.e. without any oscillatory features), $P_{\rm ref}$, is constructed 
from the measured power spectrum using a cubic spline fit over the wavenumber range 
$0.0046 < (k/h {\rm Mpc}^{-1}) < 1.2$, using the measured spectrum smoothed over 
25 bins in wavenumber. The spline is constrained to pass through the data points in 
this coarse rebinning of the measured power spectrum. 

\item[2.]
We compute the ratio, $R$(k), of the measured power spectrum, $P(k)$, to the reference spectrum, 
$P_{\rm ref}(k)$,  obtained in step 1: 

\begin{equation}
R(k) = \frac{P(k)}{P_{ \rm ref}(k)}.
\end{equation}

\item[3.] 
A linear perturbation theory power spectrum is generated with {\tt CAMB} for the cosmological 
parameters used in the {\tt BASICC} simulation. A smooth reference spectrum, 
$P^{\rm L}_{\rm ref}$, is defined for this spectrum in the same manner as described for the 
measured spectrum in Step 1, using the same wavenumber bins. A ratio, $R_{\rm L}$, 
is derived for the linear perturbation theory spectrum by dividing by this reference spectrum. 

\item[4.]
The linear theory ratio, $R_{\rm L}$, is compared with the measured ratio, $R$. Two modifications 
are considered to the linear theory ratio. The first is a stretch or scaling of the wavenumber used in 
the linear theory ratio, as described above, to mimic the act of changing the dark energy equation 
of state parameter, $w$. The goal here is to see what variation in $w$ can be tolerated before $R_{\rm L}$ 
is no longer a good fit to the measured ratio $R$. The second change is to allow for a damping of the 
oscillations beyond some characteristic wavenumber by multiplying the theoretical power spectrum by 
a Gaussian filter:  
\begin{equation}
W(k) = \exp \left(- \frac{k^2}{2 k_{\rm nl}^2} \right),
\end{equation}
where $k_{\rm nl}$ is a free parameter. 
Hence, the linear theory ratio is modified to: 
\begin{equation}\label{eq:nl}
R_{\rm L}(k) =  \left( \frac{P^{L}}{P^{L}_{ \rm ref}}  (\alpha k) - 1 \right)\times  W(k,k_{\rm nl})+ 1 
\end{equation} 

\item[5.] 

A likelihood is computed for each combination of the parameters $k_{\rm nl}$ and $\alpha$, assuming 
Gaussian errors: 
\begin{equation}\label{eq:lik}
- 2 \ln L = \chi^2 = \sum_{i} \left( \frac{ R^{i} - R_{\rm L}^{i} }{\sigma^{i}/P^{i}} \right)^2
\end{equation}
where the summation is over wavenumber and $\sigma^{i}$ is the error on the power spectrum estimated 
in the \textit{i}$^{\rm th}$ bin (as given by Eq.~\ref{eq:err}).  
We generate a grid of models using $200^2$ different combinations of $\alpha$ and $k_{\rm nl}$ in the 
ranges [0.9,1.1] and [0,0.4] respectively.

\item[6.]
Finally, the best fit values for $\alpha$ and $k_{\rm nl}$ correspond to 
those for the model with the maximum likehood. 
We obtain confidence limits on the parameter estimation  
by considering the models within $\Delta \chi^2$ equal to $2.3$ and $6.0$; 
in the case of a Gaussian likelihood, these would correspond to  
the $68\%$ ($1$-$\sigma$ error) and $95\%$ ($2$-$\sigma$ error) 
confidence levels on the best fit. We note that in some cases presented 
later (see Fig. 15), the distribution of the likelihood is not Gaussian. 

\end{enumerate}
In some cases, we also present constraints on $w$ derived using a slightly modified version of 
the approach of Blake \& Glazebrook. The main difference is that we follow step 1 to construct 
a ratio from the measured power spectrum, rather than using a zero-baryon transfer function.

One issue to be resolved is the range of wavenumbers which should be used in the fitting process. 
To address this, we used the power spectrum of the dark matter measured at $z=6$. We systematically 
varied the minimum and maximum wavenumbers used in our fit and compared the values of the scaling 
parameter, $\alpha$, recovered. Our results are fairly insensitive to the choice of the maximum 
wavenumber, particularly when damping of the oscillations is included in the fitting algorithm. 
However, the recovered $\alpha$ shows a systematic shift once the minimum wavenumber exceeds 
$k \sim 0.1 h {\rm Mpc}^{-1}$. For minimum wavenumbers smaller than this, there is little difference 
in the recovered value of $\alpha$ or in the size of the errors on $\alpha$, as these modes have 
relatively large errors in our simulation. This is encouraging news for realistic survey geometries, 
for which the power spectrum measured at low wavenumbers will be distorted due to the window 
function of the survey. In the rest of the paper, we use the power spectrum in the wavenumber 
interval $k/(h{\rm Mpc}^{-1})=[0, 0.4]$ to constrain the value of $\alpha$. 

\section{Results}

\begin{figure}
\includegraphics[width=8.5cm]{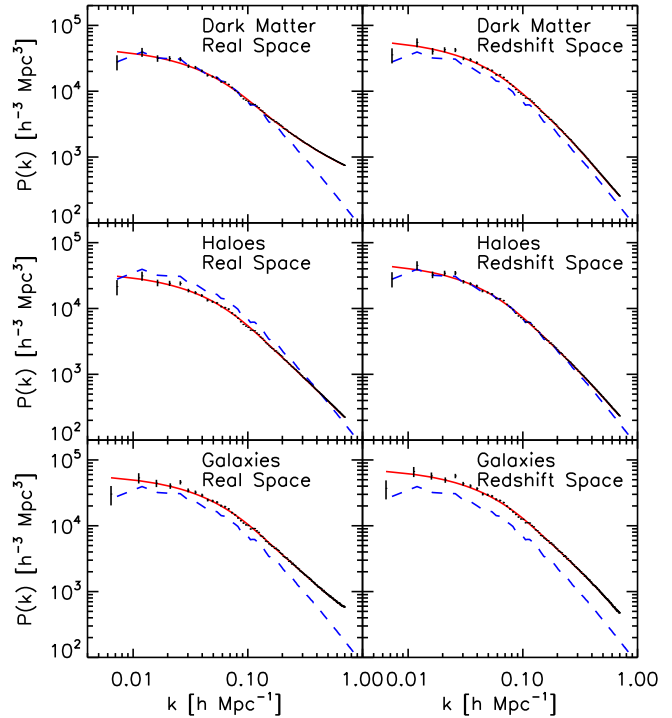}
\caption{ 
The power spectra of dark matter particles, dark matter haloes and galaxies at $z=0$ (error bars). 
The real-space power spectra are plotted in the left hand column and the redshift-space power spectra 
appear in the right hand column. The red curves show the reference spectra derived from the measured 
spectra using a cubic spline fit, as described in Section 4. The blue curve is the same in each panel, 
showing the linear perturbation theory prediction for the $z=0$ matter power spectrum (plotted using a 
high redshift output obtained from the {\tt BASICC} simulation, which has been scaled in amplitude according to the difference in growth factors between the two epochs expected in linear perturbation theory)
The errors on the power spectrum are estimated using Eq.~\ref{eq:err}.  
\label{fig:fit1}}
\end{figure}

\begin{figure}
\includegraphics[width=8.5cm]{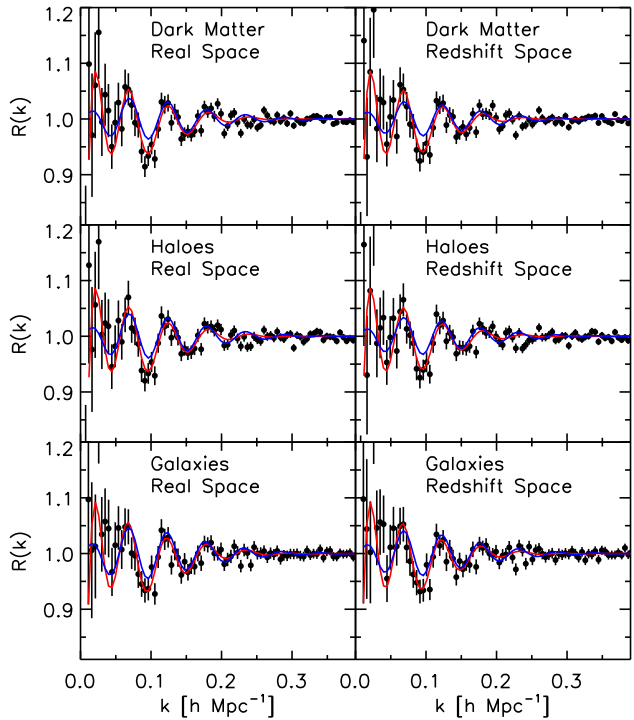}
\caption{ 
The ratio of the measured power spectrum divided by a smooth reference spectrum. 
The symbols correspond to the measurements plotted in Fig.~\ref{fig:fit1} divided 
by the red curve in each panel of that figure. The red lines here show the 
best-fitting model in each case using the general method and the blue curves 
show the best fit for the parametric method. The errors on the power 
spectrum are estimated 
using Eq.~\ref{eq:err}.}  
\label{fig:fit2}
\end{figure}

In this Section, we present the expected constraints on the dark matter equation of state using 
the power spectra measured from our simulations. We first show how our algorithm for extracting 
the equation of state parameter works in practice, for dark matter particles, haloes and galaxies,  
comparing the results obtained in real-space and redshift-space (\S 5.1). 
We then assess the need for an accurate model of the linear theory power spectrum and the 
relative merits of the general and parametric fitting procedures (\S 5.2).    
In \S 5.3, we present our main results, which are summarized in Fig.~\ref{fig:res} 
and Table~\ref{tab:gals}, which lists the best-fitting value of $\alpha$ and the estimated 
error for different samples of galaxies at $z=1$, along with the corresponding fractional error in $w$. 
Finally, in \S 5.4, we use the results presented in \S 5.3 to make forecasts for the 
accuracy with which several forthcoming surveys will be able to measure the value of $w$.

\subsection{The algorithm to extract the scale of the acoustic oscillations in action} 
We present a series of plots for samples at $z=0$, which illustrate the 
various stages in the fitting process. Fig.~\ref{fig:fit1} shows the power spectra measured for 
different tracers, both in real-space and redshift-space. The sample of dark matter haloes includes 
all objects with a mass in excess of $5.4 \times 10^{12}\,h^{-1}\,M_{\odot}$. The galaxy sample 
is magnitude-limited with a space density of $\bar{n}=5\times10^{-4}\,h^{-3}\,{\rm Mpc}^3$. For reference, 
the linear perturbation theory 
power spectrum for the mass at $z=0$ is shown by the blue line in each panel: this is the power 
spectrum of the dark matter measured in real-space at $z=15$, scaled by the ratio of growth factors 
in order to have the amplitude expected at $z=0$. It is important to bear in mind 
that the y-axis in this plot covers more than 
a factor of one thousand in amplitude. Fig.~\ref{fig:fit1} shows that there is considerable 
variation in the power spectra measured for different types of objects, and between 
the results in real-space and redshift-space, which re-inforces the points made in Section 3 regarding 
deviations from the predictions of linear perturbation theory on large scales. The red curve 
in each panel shows the corresponding reference power spectrum, which is constructed from the 
measured power spectrum as explained in Section~4. 

\begin{figure}
\includegraphics[width=8.5cm]{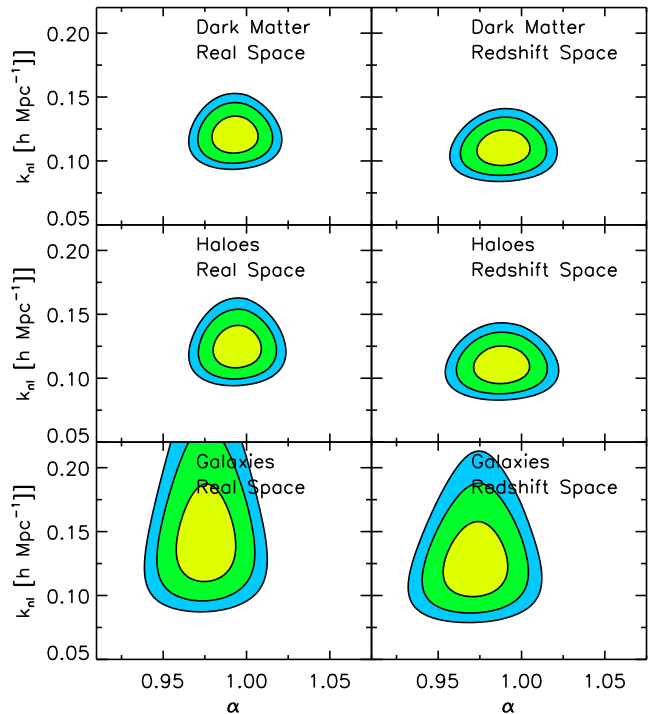}
\caption{ The constraints on the parameters $k_{\rm nl}$ and $\alpha$ for the power spectra plotted in 
Fig.~\ref{fig:fit1}. The contours show the 1, 2 and $3$-$\sigma$ confidence 
limits for two parameters. 
\label{fig:fit3}}
\end{figure}

In Fig.~\ref{fig:fit2}, the symbols show the ratio obtained by dividing the measured power spectrum 
by the appropriate reference spectrum for the same samples plotted in Fig.~\ref{fig:fit1}. The ratios look remarkably similar for the different tracers up 
to $k \approx 0.15 h {\rm Mpc}^{-1}$. Beyond this wavenumber, the appearance of the oscillations varies 
from panel to panel, but the ratio stays close to unity. This similarity illustrates how well the 
approach for producing the reference spectrum works. The red curves in each panel show the best-fitting 
model produced in the general scheme whilst the blue curves show the fit obtained in the parametric approach. 
The best fits have somewhat different forms at wavenumbers below $ k \sim 0.05 h {\rm Mpc}^{-1}$. 
The constraints on the values of the parameters $k_{\rm nl}$ and $\alpha$ are presented in Fig.~\ref{fig:fit3}, 
where we show the 1, 2 and 3-$\sigma$ ranges in the case of two parameters, computed assuming Gaussian 
errors. There is a weak systematic trend for the best-fitting result for $\alpha$ to shift to lower 
values when galaxies are considered instead of the dark matter. The errors on the recovered parameters are 
larger in the case of galaxies than for the dark matter or for haloes, reflecting the lower signal-to-noise 
of the predicted galaxy power spectrum.  

\begin{figure}
\includegraphics[width=8.5cm]{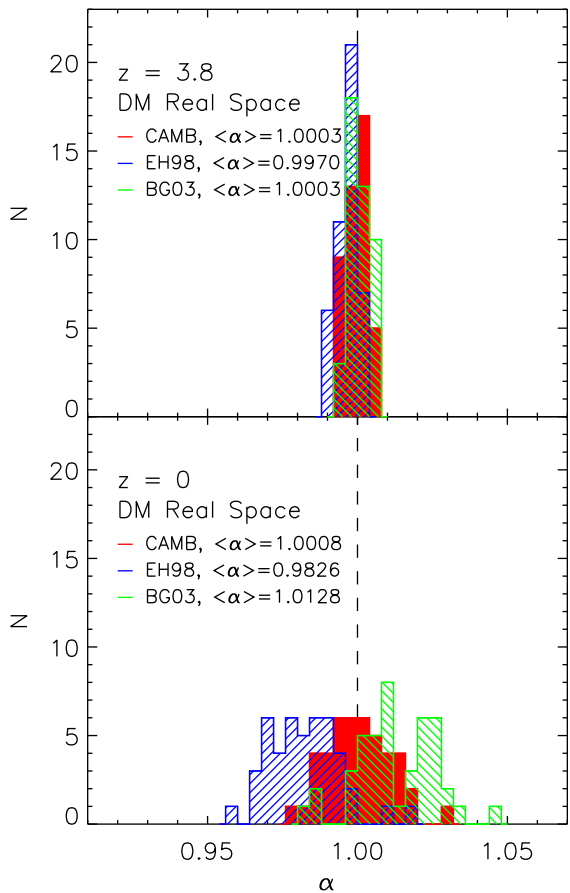}
\caption{ 
The best fit value for the scaling parameter $\alpha$, recovered from the ensemble 
of low resolution simulations, using the dark matter power spectrum in real-space. 
The results are show for two different redshifts: $z=3.8$ (top) and $z=0$ bottom. 
The histograms marked {\tt CAMB} and BG03 show the results for the general and 
parametric fitting procedures, respectively. The blue histogram shows the results 
if the general method is followed with the {\tt CAMB} power spectrum replaced by the 
formula for the linear theory power spectrum presented by Eisenstein \& Hu (1998). 
\label{fig:hist}}
\end{figure}

\subsection{Two tests of the algorithm}

Before presenting the main results of applying our algorithm to extract the 
acoustic oscillation scales for various samples drawn from the {\tt BASICC} run, 
we use the {\tt L-BASICC} ensemble to address two questions: 
1) How accurately do we need to model the linear perturbation theory matter power 
spectrum to avoid introducing a systematic bias into the results for the oscillation 
scale? 
2) How does the performance of the new method for constraining the oscillation scale 
introduced in this paper compare with earlier approaches? 
To help answer these questions, we use the power spectrum of the dark matter 
measured from the {\tt L-BASICC} runs in real-space at $z=0$ and $z=3.8$, the 
highest output redshift besides the initial conditions. 
The results of applying our standard algorithm for extracting the oscillation 
scale are shown by the red histogram labelled {\tt CAMB} in Fig.~\ref{fig:hist}, 
which gives the distribution of the best-fitting value of $\alpha$. 
The ensemble returns an unbiased mean value for the stretch parameter, $\alpha=1$. 
At $z=3.8$, the standard deviation on the best fit is 0.3\%; by $z=0$, this rises 
to $1\%$. 

To address the first issue above, regarding how well we need to model the linear theory 
power spectrum to get an unbiased result for the oscillation scale, we replace 
the {\tt CAMB} generated power spectrum in our algorithm by the approximation 
introduced by \cite{EH1998}. These authors proposed a physically motivated 
expression for the linear theory power spectrum, with parameters set 
to achieve a reasonable match to the results obtained from detailed calculations 
using Boltzmann codes over a much wider range of wavenumbers than are typically 
considered for baryonic acoustic oscillations. Eisenstein \& Hu's motivation 
was to provide physical insight into the form of the power spectrum in a cold 
dark matter universe and to produce a code which could rapidly calculate large 
numbers of power spectra for grids cosmological parameters. Of course, the correct 
approach in our fitting procedure is to use the same code to compute the linear theory 
spectrum as was used to generate the initial conditions in the N-body simulation. 
In the case of real data, we do not have the luxury of knowing which Boltzmann code to 
use, so we should use the one which claims to be the most accurate representation 
of the model we are testing. Nevertheless, it is instructive to perform this test to 
see what error is introduced by using a less accurate calculation of the transfer 
function. The choice of Eisenstein \& Hu's code is particularly relevant for this 
purpose as Blake \& Glazebrook used this formalism to inspire their parametric 
expression to fit the acoustic oscillations. The use of Eisenstein \& Hu's 
formalism to model the linear theory power spectra generated with {\tt CAMB} 
introduces a small but measurable systematic shift in the mean value of $\alpha$. 
At $z=0$, the mean $\alpha$ indicated by the blue histogram in Fig.~\ref{fig:hist} 
is $0.98 \pm 0.01$. 

We answer the second question by adopting the fitting algorithm of Blake \& Glazebrook 
(2003), which assumes a parametric form for the ratio of the power spectrum with 
baryons to a smooth, cold dark matter only power spectrum. Changing the fitting 
method in this way also introduces a similar magnitude of shift in the best-fitting 
value of $\alpha$. The green histogram shows the results when we use the parametric 
approach introduced by Blake \& Glazebrook (2003). The mean value of $\alpha$ in this 
case is $1.01 \pm 0.01$. These shifts are small but one must bear in mind that the 
corresponding bias in the dark energy equation of state parameter is several times 
larger than the shift in $\alpha$.

\begin{table*}
\begin{center}
\begin{tabular}{ccc|cccccc|cccccc}
	 \mc{15}{c}{$z=0$} \\
\hline
\hline
         &  Sel I & Sel II  & \mc{6}{c}{Real-space}          & \mc{6}{c}{Redshift-space}  \\ 

 id      &  $\bar{n}$                                                     & & 
$b$ & $\bar{n}P$ &  $k_{\rm nl }       $ & $\alpha$ & $ \Delta \alpha$ & $\Delta \alpha $ &  
$b$ & $\bar{n}P$ &  $k_{\rm nl }       $ & $\alpha$ & $ \Delta \alpha$ & $\Delta \alpha $    \\
         &  $ h^{3}{\rm Mpc}^{-3}$                                   & &   
    &            & $h/{\rm Mpc}$  &          &        \%      &   \%    & 
    &            & $h/{\rm Mpc}$  &          &        \%      &   \%       \\   
         &                                     & &   
    &            &   &          &              &   (SE07)    & 
    &            &   &          &              &   (SE07)       \\   
\hline

  DM & $                $ &                   & 0.99 & 3567 & 0.120 & 0.993 & 0.91 & 1.02            & 1.15 & 3635 & 0.110 & 0.989 & 1.05 & 1.17\\
  A  & $5.0e-4$ &                   & 1.18 & 1.78 & 0.144 & 0.975 & 1.16 & 1.10            & 1.32 & 2.15 & 0.125 & 0.972 & 1.26 & 1.23\\
  B  & $2.5e-4$ &                   & 1.33 & 1.11 & 0.155 & 0.971 & 1.34 & 1.18            & 1.47 & 1.34 & 0.139 & 0.966 & 1.35 & 1.23\\
  C  & $2.5e-4$ & red               & 1.32 & 1.15 & 0.152 & 0.978 & 1.35 & 1.21            & 1.46 & 1.36 & 0.127 & 0.975 & 1.49 & 1.37\\
  D  & $2.5e-4$ & strong            & 1.06 & 0.67 & 0.155 & 0.956 & 1.75 & 1.41            & 1.20 & 0.86 & 0.138 & 0.956 & 1.67 & 1.42\\
  E  & $2.5e-4$ & blue              & 1.03 & 0.66 & 0.141 & 0.964 & 1.92 & 1.56            & 1.17 & 0.83 & 0.130 & 0.962 & 1.79 & 1.53\\
  F  & $2.5e-4$ & weak              & 1.30 & 1.16 & 0.132 & 0.980 & 1.55 & 1.40            & 1.44 & 1.34 & 0.115 & 0.972 & 1.66 & 1.54\\
 haloes & $5.9e-5$ &                   & 1.56 & 0.81 & 0.197 & 0.980 & 1.32 & 1.07            & 1.71 & 1.04 & 0.148 & 0.975 & 1.43 & 1.25\\

\hline
\end{tabular}
\\
\begin{tabular}{ccc|cccccc|cccccc}
\\
	 \mc{15}{c}{$z=1$} \\
\hline
\hline
         &  Sel I & Sel II  & \mc{5}{c}{Real-space}          & \mc{5}{c}{Redshift-space}  \\ 

 id      &  $\bar{n}$                                                     & & 
$b$ & $\bar{n}P$ &  $k_{\rm nl }       $ & $\alpha$ & $ \Delta \alpha$ & $\Delta \alpha $ &  
$b$ & $\bar{n}P$ &  $k_{\rm nl }       $ & $\alpha$ & $ \Delta \alpha$ & $\Delta \alpha $    \\
         &  $ h^{3}{\rm Mpc}^{-3}$                                   & &   
    &            & $h/{\rm Mpc}$  &          &        \%      &   \%    & 
    &            & $h/{\rm Mpc}$  &          &        \%      &   \%       \\   
         &                                     & &   
    &            &   &          &              &   (SE07)    & 
    &            &   &          &              &   (SE07)       \\   
\hline
  DM & $                $ &                    & 0.99 & 1269 & 0.163 & 0.997 & 0.61 & 0.68            & 1.29 & 1710 & 0.133 & 0.991 & 0.77 & 0.88\\
  A  & $5.0e-4$ &                    & 1.34 & 0.87 & 0.188 & 0.980 & 1.30 & 1.10            & 1.60 & 1.19 & 0.164 & 0.976 & 1.21 & 1.07\\
  B  & $2.5e-4$ &                    & 1.31 & 0.43 & 0.212 & 0.975 & 2.02 & 1.47            & 1.57 & 0.59 & 0.174 & 0.970 & 1.72 & 1.38\\
  C  & $2.5e-4$ & red                & 1.39 & 0.48 & 0.235 & 0.977 & 1.81 & 1.32            & 1.65 & 0.65 & 0.208 & 0.975 & 1.52 & 1.17\\
  D  & $2.5e-4$ & strong             & 1.31 & 0.40 & 0.624 & 0.971 & 1.90 & 1.14            & 1.57 & 0.55 & 0.186 & 0.970 & 1.79 & 1.31\\
  E  & $2.5e-4$ & blue               & 1.30 & 0.40 & 0.219 & 0.973 & 2.31 & 1.47            & 1.56 & 0.54 & 0.159 & 0.962 & 1.98 & 1.48\\
  F  & $2.5e-4$ & weak               & 1.37 & 0.47 & 0.218 & 0.987 & 1.91 & 1.38            & 1.63 & 0.64 & 0.190 & 0.978 & 1.61 & 1.25\\
 haloes & $5.9e-5$ &                    & 3.07 & 0.59 & 0.226 & 1.000 & 1.65 & 1.24            & 3.34 & 0.77 & 0.146 & 0.994 & 1.82 & 1.53\\

\hline
\end{tabular}

\caption{
The results of applying the general fitting procedure described in \S 4 to  
 power spectra measured for different galaxy catalogues at $z=0$ (top) and $z=1$ (bottom). 
In each table, the first row gives the results for the dark matter and the final row 
lists results for a sample of dark matter haloes (all haloes with mass in excess 
of $2.7 \times 10^{13}\,h^{-1}\,M_{\odot}$). The first column gives the label of the sample, 
as defined in Section 2. The second column gives the space density of galaxies. The first 
two samples, A and B, are constructed by applying a magnitude limit. Samples C-F are derived 
from sample A by applying a second selection criterion,  as listed in the third column. 
Samples C and E correspond to the red and blue halves of sample A respectively. Samples D 
and F comprise the $50\%$ of galaxies from sample A with the strongest and weakest 
(in terms of equivalent width) OII[3727] emission lines, respectively. 
Column 4(10) gives the effective bias of the sample, computed from the square root of the ratio of the 
measured galaxy power spectrum in real (redshift) space to the real space power spectrum of the dark matter over the wavenumber 
interval $0.01 < (k/h {\rm Mpc}^{-1}) < 0.05$. 
Column 5(1`) gives the ratio of the clustering signal to the shot noise for the power spectrum measurement, averaged over 
the wavenumber range $0.19 < (k/h {\rm Mpc}^{-1}) < 0.21$.
Columns 6 and 7 (12 and 13) give the best-fitting values of the scaling parameter $\alpha$ and the $1-\sigma$ error on the fit, in real (redshift) space. 
Column 9 (15) gives the error expected on the scale parameter from 
Seo \& Eisenstein (2007). The rms lagrangian displacement was set equal
to 1 over the best fit non linear scale ($1/k_{nl}$) for each case.  
}
\label{tab:gals}

\end{center}
\end{table*}

dsf

\begin{figure}
\includegraphics[width=8.5cm]{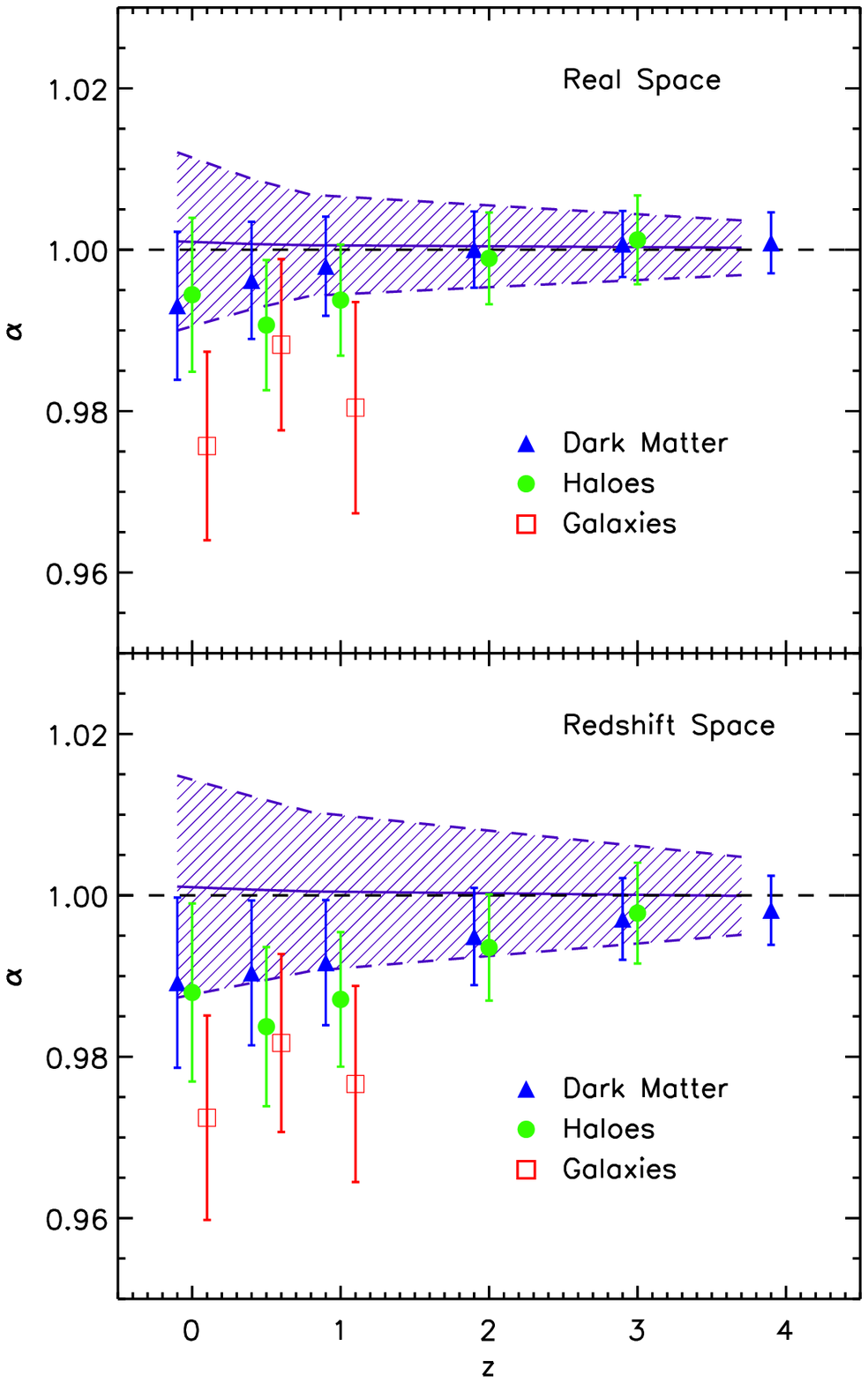}
\caption{ 
The best-fitting value of the scale factor $\alpha$ as a function of redshift, for different 
tracers of the density distribution, in real-space (top) and redshift-space (bottom). The symbols 
show results from the high resolution {\tt BASICC} simulation: dark matter (blue triangles), dark matter 
haloes with mass in excess of $5.4\times10^{12}\,h^{-1}\,M_{\odot}$ (green circles) and galaxies 
(red squares). The error bars show the 1-$\sigma$ range on $\alpha$, calculated from $\Delta \chi^2$. 
The hatched region shows 
the central $68\%$ range of the results obtained using the dark matter in the ensemble of low 
resolution simulations. Recall that $\alpha=1$ corresponds to an unbiased measurement of the 
equation of state parameter, $w$, and that $\delta w \approx 4 \delta \alpha$ at $z=1$. 
\label{fig:alpha}
}
\end{figure}

\begin{figure}
\includegraphics[width=8.5cm]{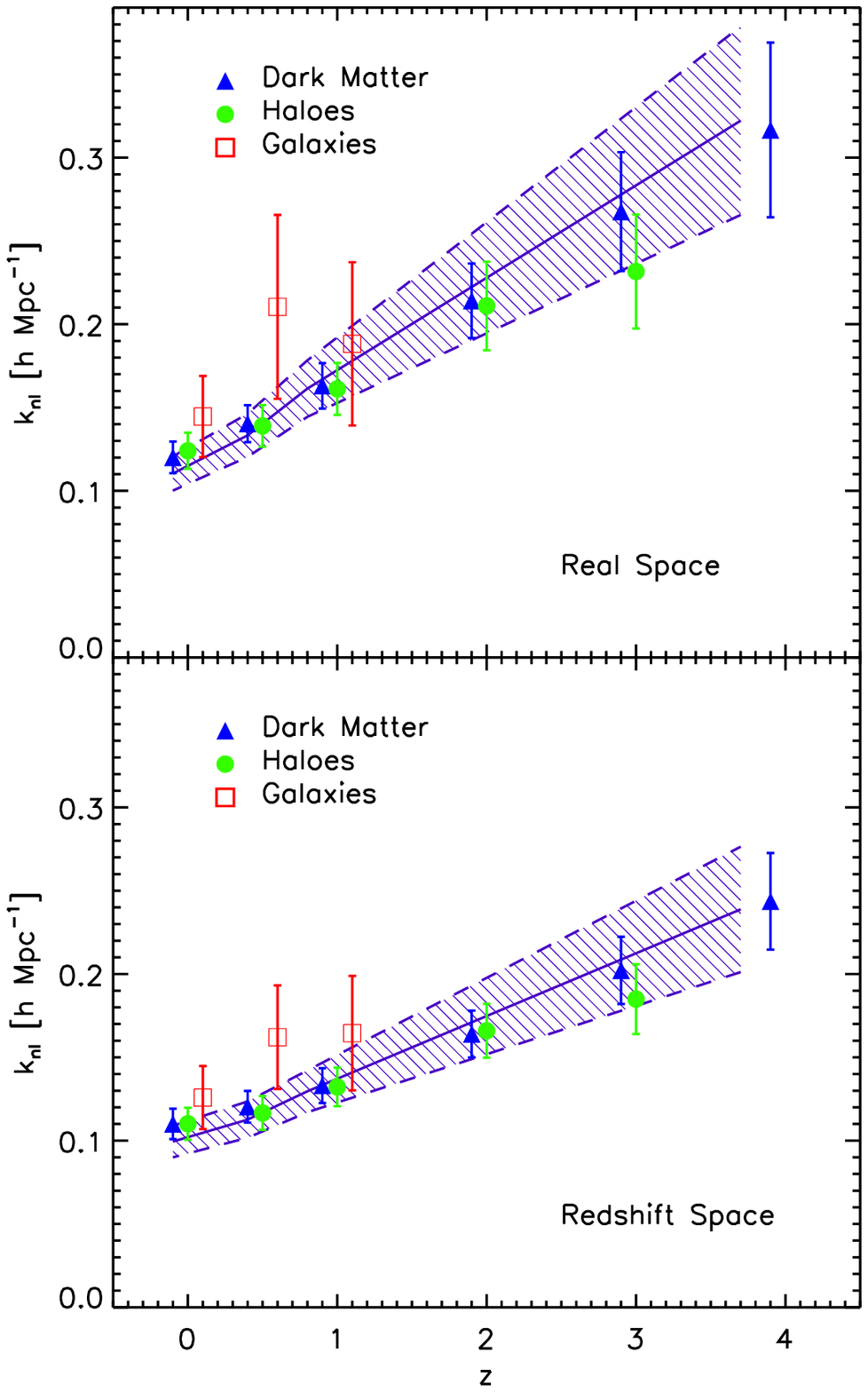}
\caption{ 
The best-fitting value of the damping scale $k_{\rm nl}$ as a function of redshift, for different 
tracers of the density distribution, in real-space (top) and redshift-space (bottom). The symbols 
show results from the high resolution {\tt BASICC} simulation: dark matter (blue triangles), dark matter 
haloes with mass in excess of $5.4\times10^{12}\,h^{-1}\,M_{\odot}$ (green circles) and galaxies 
(red squares). The error bars show the 1-$\sigma$ range on $\alpha$. The hatched region shows 
the central $68\%$ range of the results obtained using the dark matter in the ensemble of low 
resolution simulations. 
\label{fig:knl}}
\end{figure}

\begin{figure}
 \includegraphics[width=8.5cm]{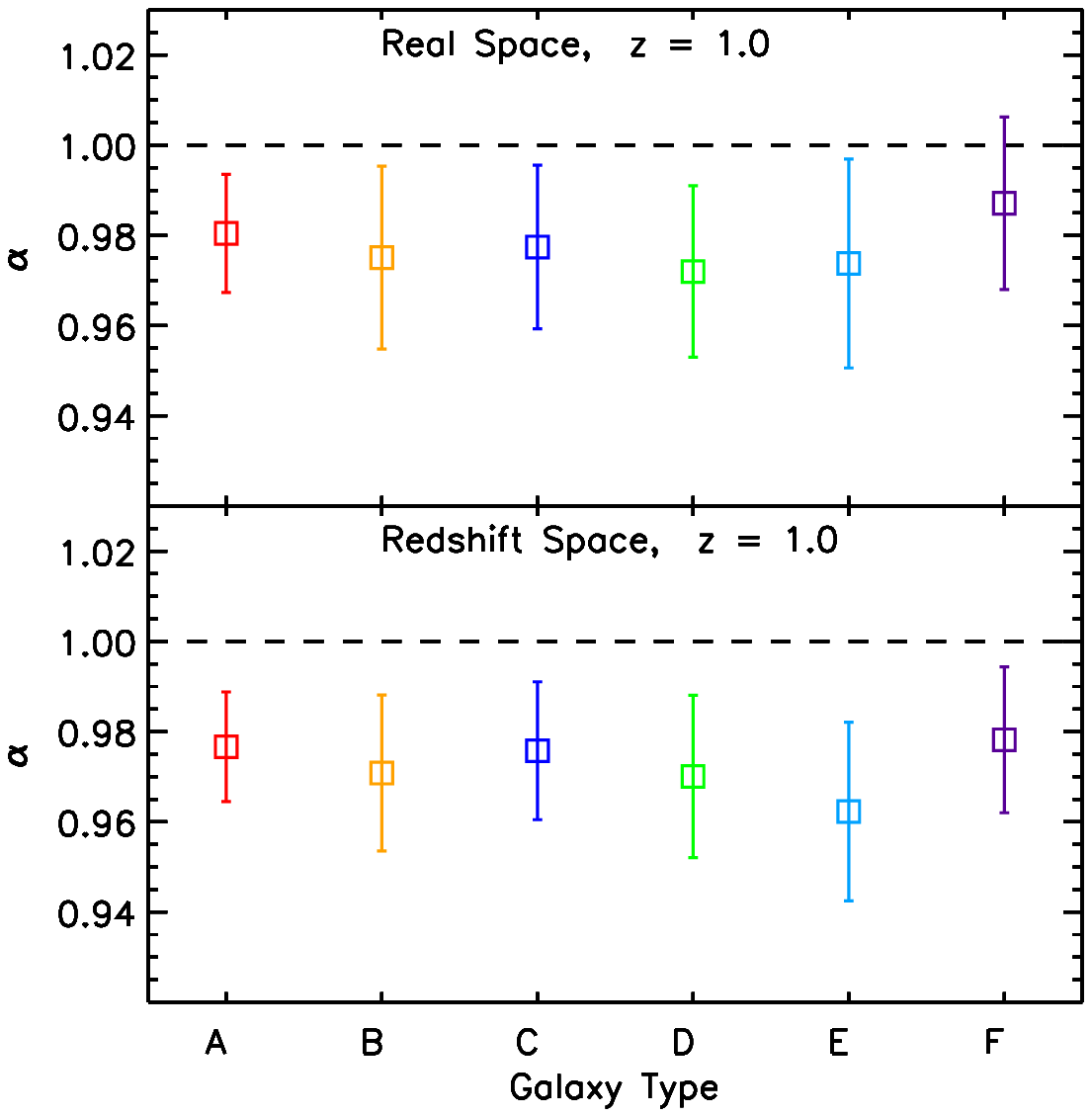}
\caption{
The recovered value of the stretch parameter $\alpha$ for the galaxy samples listed in Table 2. 
Recall that $\alpha=1$ corresponds to the equation of state parameter $w=-1$. At $z=1$, a shift 
in $\alpha$ away from unity implies a shift in the recovered value of $w$ given by 
$\delta w \approx 4 \delta \alpha$.}
\label{fig:res}
\end{figure}

\subsection{The main results} 
We now turn our attention back to the general results shown in Fig.~\ref{fig:alpha} 
and Fig.~\ref{fig:knl}, and discuss the conclusions for different tracers of the density 
field in turn. In these plots, the symbols refer to the constraints obtained from the high 
resolution {\tt BASICC} simulation and the shading shows results from the ensemble of low resolution 
simulations, {\tt L-BASICC}. 

The blue triangles in Fig.~\ref{fig:alpha} show the values obtained for $\alpha$ from 
the power spectrum of the dark matter. There is a trend for the best-fitting value to 
deviate away from unity with decreasing redshift, although the result at $z=0$ is still 
within 1-$\sigma$ of $\alpha=1$. The mean of the ensemble of low resolution runs does not, 
however, show any deviation away from $\alpha=1$ as a function of redshift, although the 
scatter on the recovered value of $\alpha$ increases towards the present day. 
If we examine the analogous results for individual simulations taken from the low resolution ensemble, 
we find a wide range of behaviour for the best-fitting value of $\alpha$ for the dark matter. 
Some low resolution runs give results which look like the 
high resolution one, whereas others show deviations away from $\alpha=1$, with values of 
$\alpha > 1$, as $z=0$ is approached. The trend seen for the dark matter in the high resolution 
run serves to illustrate the importance of sampling fluctuations, even in such large volumes. 
In redshift-space, the scatter in the recovered value of $\alpha$ 
is larger than in real-space (see also \citealt{Seo2005,Eisenstein2006}). 

To obtain the errors quoted in Table 2 on the parameters $\alpha$ and $k_{\rm nl}$, we 
assume Gaussian mode counting errors on the power spectra measured in the {\tt BASICC} simulation, 
as given by Eq. 3. In Fig. 3, we showed that this simple estimate of the errors on the power 
spectrum agreed fairly well with the scatter found in the measurements from the {\tt L-BASICC} 
ensemble, particularly for the case of the dark matter. We have extended this comparison to 
look at how the errors on $\alpha$ and $k_{\rm nl}$ quoted in Table 2 match the scatter in 
these parameters obtained from the {\tt L-BASICC} runs. We find the scatter estimated from the 
ensemble is somewhat larger than the error inferred using the mode counting argument. At $z=0$, 
the mode counting errors are $20\%$ smaller for $\alpha$ for the dark matter in real space. 
In redshift-space, the discrepancy increases to nearly $30\%$. The mismatch between the two estimates 
is smaller at $z=1$. The level of disagreement is not remarkable. It could be the case that the 
scatter from the ensemble has not converged, even with 50 realizations of the density field. 
A more likely explanation, particularly in view of the redshift dependence of the discrepancy, 
is mode coupling in the power spectrum measurements arising from nonlinearities and redshift space 
distortions, which could increase the variance in the power spectrum compared with the Gaussian estimate.  

Fig.~\ref{fig:knl} shows that there is a strong trend for the best-fitting value of the 
smoothing scale, $k_{\rm nl}$, to decrease with decreasing redshift. This results from the 
oscillations being erased and modified down to smaller wavenumbers as the nonlinearities 
in the density field grow. 
The variation of the smoothing scale $k_{\rm nl}$ on redshift 
is well described by a linear relation: $k_{\rm nl} = a + b z$. In real-space, 
$a = 0.108 \pm 0.0082$ and $b = 0.054 \pm 0.0110$. In redshift-space, $a = 0.096\pm 0.0074$ 
and $b = 0.036 \pm 0.0094$.

The constraints on $\alpha$ and $k_{\rm nl}$ for dark matter haloes (with masses in excess of 
$5\times 10^{12}\,h^{-1}\,M_{\odot}$) are plotted with green circles in  Figs.~\ref{fig:alpha} 
and ~\ref{fig:knl}. The parameter constraints obtained for this sample of haloes are very similar 
to those found for the dark matter, except for the value of $k_{\rm nl}$ at high redshift. 
Considering haloes in place of dark matter represents a step closer to the observations, so it 
is reassuring that the conclusions do not change significantly. 

Finally, in Figs.~\ref{fig:alpha} and ~\ref{fig:knl}, we show using red squares the results for 
magnitude-limited samples of galaxies. The magnitude limit is varied with redshift such that in 
each case the galaxy sample has a space density $\bar{n}=5\times10^{-4}\,h^{-3}\, {\rm Mpc}^3 $. There is 
a weak systematic shift in the best-fitting values of $\alpha$ compared with the results obtained 
for the dark matter. At the same time the signal to noise of the power spectrum measurement is 
lower for the galaxy samples than for the dark matter, so the errors on the best-fitting parameters 
are correspondingly larger for the galaxies. The galaxy samples are consistent with $\alpha=1$ at 
slightly over 1-$\sigma$. The size of this systematic shift is comparable to the random measurement 
errors, so we cannot reach a firm conclusion. It will be very interesting to repeat our calculation with 
a larger simulation volume to reduce the size of the random errors and to assess if such shifts 
could genuinely provide an ultimate limitation to the accuracy of this method. 

As a result of using a semi-analytic galaxy formation model  which makes
predictions for the observable properties of galaxies, we can vary the
selection criteria used to construct samples and compare the
constraints on the equation of state. The results of this exercise at $z=1$ are
presented in Table~2 and in Fig.~\ref{fig:res}, where we consider a
range of samples defined either by a simple magnitude limit, or by a
magnitude limit applied in combination with a colour cut or
a restriction on the strength of an emission line. The key result from
comparing the constraints for different samples is that whilst there
are no strong systematic differences between the results, the accuracy
of the constraints varies significantly. For example, using a
catalogue of red galaxies, we predict that one could measure the dark
energy equation of state with an accuracy $~40\%$ better than with the
same number density of galaxies chosen by the strength of their emission
lines.

We compare the error on the acoustic scale extracted from our simulations 
with the results of the prescription set out by \cite{Seo2007}. 
The \cite{Seo2007} algorithm contains a parameter which is equivalent 
to $1/k_{\rm nl}$. If we use our best fitting values of $k_{\rm nl}$, 
we find that the Seo \& Eisenstein prescription gives a similar estimate 
of the error on the acoustic scale to that we obtain by fitting directly 
to the simulation results. However, if we use the value of $k_{\rm nl}$ 
suggested by \cite{Seo2007}, which they extract from a dark matter simulation, we find that their prescription gives an optimistic estimate of the error 
on $\alpha$. The reason we recover a larger value of $k_{\rm nl}$ from 
our galaxy samples than we do for the dark matter is due to the increased 
discreteness shot noise in these samples, which results in noisier power 
spectra at high $k$. This causes an elongation in the confidence levels 
in the $k_{\rm nl}$ versus $\alpha$ plane.  

It is interesting to compare the results for the dark matter and for 
the galaxy samples with those for a set of massive haloes. Table 2 also 
gives the constraints on $\alpha$ and $k_{\rm nl}$ for a sample of 
massive haloes (see also Angulo et~al. 2005). 
There are 142\,000 haloes in the {\tt BASICC} output at z=1 with a 
mass in excess of $2.7 \times 10^{13}\,h^{-1}\,M_{\odot}$. 
Although the effective bias of this sample of massive haloes is greater 
than that of any of the galaxy samples listed in Table~2, the 
reduction in space density means that $\bar{n}P \approx 1$ and the 
estimated error on $w$ is comparable to that found for the galaxy samples.

\subsection{Forecasts for future surveys}

We can use the results presented in Table~2 to make a rough estimate of 
the accuracy with which future surveys are likely to be able to constrain the 
scale of the acoustic oscillations.
This can be done using a simple calculation motivated by the expression 
for the fractional error in the power spectrum given by Eq. 3.  
We assume that the error in the distance scale, $\Delta \alpha$, scales with 
the volume of the survey, $V_{\rm survey}$, and the product of the space density of 
galaxies and the power spectrum, $\bar{n}P(k=0.2h{\rm Mpc}^{-1})$, as: 

\begin{equation}
\Delta \alpha \propto \frac{1}{\sqrt{V_{\rm survey}}} \left( 1 + \frac{1}{\bar{n}P}\right).
\end{equation} 
The constant of proportionality can be set for a particular galaxy sample using the results 
given in Table 2. 

The WiggleZ survey is currently underway and will measure redshifts for 400,000 
blue galaxies over 1000 square degrees in the redshift interval $z=0.5-1.0$ (Glazebrook et~al. 2007).
For the cosmological parameters adopted in this paper, this gives a comoving volume of 
$1.13 \,h^{-3}\,{\rm Gpc}^{3}$. Using the blue colour selected sample or the large equivalent 
width sample from Table 2, and assuming $\bar{n}P \sim 1$ for WiggleZ galaxies, somewhat higher 
than we find in our simulation, we estimate that this survey will measure the distance scale to 
an accuracy of $\Delta \alpha \sim 2\% $, which is similar to 
that claimed by Glazebrook et~al. using linear perturbation theory.

The WFMOS survey has been proposed to motivate the construction of a new spectrograph 
for the Subaru telescope (Glazebrook et~al. 2005). This will target galaxies with a space density 
of $\bar{n} = 5 \times 10^{-4} \,h^{3}\,{\rm Mpc}^{-3}$ in the redshift interval $z=0.5-1.3$ 
over 2000 square degrees, covering a volume of $4.4 \,h^{-3}\, {\rm Gpc}^{3}$. (There is also 
a WFMOS survey which will target $z=3$ galaxies but over a much smaller solid angle.) Using 
sample A from Table~2, and adopting $\bar{n}P = 1$, we obtain an estimated error 
of $\Delta \alpha = 0.83\% $, again in good agreement with Glazebrook et~al. 

Photometric surveys can generally cover a larger solid angle than spectroscopic surveys 
down to a fainter magnitude limit. The fainter magnitude limit results in a higher median 
redshift and a broader redshift distribution for the survey galaxies, which means that a 
larger volume is covered. However, the limited accuracy of photometric redshift estimates 
means that in practice Fourier modes are lost and the effective volume of the survey is 
greatly reduced. Blake \& Bridle (2005) estimate that the factor by which the survey volume is 
reduced is $\approx 12 \left( \delta z /(1+z)/0.03 \right)$, where $\delta z / (1+z)$ is 
the error in the photometric redshifts. 

The Panoramic Survey Telescope and Rapid 
Response System (Pan-STARRS) survey will map $3 \pi$ steradians of the sky 
(http://pan-starrs.ifa.hawaii.edu/public/home.html). Cai et al. (2007, in preparation) show 
that the median redshift of the $3 \pi$ survey will be $z \approx 0.5$, with a tail extending 
to $z \approx 1.2$. The volume of the survey, assuming that $20,000$ square degrees cover 
low-extinction parts of the sky and give high quality clustering measurements, is 
around $41 \,h^{-3}\,{\rm Gpc}^{3}$. Talking sample A from Table 2, and setting $\bar{n}P >> 1$, 
as appropriate for the relatively high space density of galaxies in a photometric sample, and 
allowing for the reduction in the effective volume caused by a photometric redshift error 
of $\delta z/(1+z) = 0.03$, gives a forecast error on the oscillation scale 
of $\Delta \alpha \sim 0.5\%$. In the more likely event that the photometric redshift errors 
are twice as large, $\delta z/(1+z) \sim 0.06$, this figure increases to 
$\Delta \alpha \sim 0.7\% $.   

Remembering the crude conversion $\Delta w \approx 4 \Delta \alpha$ from 
Section 4, this means that the next generation of galaxy surveys is unlikely 
to deliver 1\% errors on a constant equation of state from BAO measurements 
used in isolation from other cosmological data. A survey with almost an order 
of magnitude more effective volume than Pan-STARRS will be needed 
to achieve this target. This will require an all-sky, spectroscopic galaxy 
redshift survey, such as the {\tt SPACE} mission being proposed to ESA's 
Cosmic Vision call. {\tt SPACE} will measure redshifts for galaxies in the 
interval $0.5 < z < 2$, covering around $150 h^{-3}{\rm Gpc}^{3}$. 
Extrapolating from Sample A, we forecast that an error in the oscillation 
scale of $\Delta \alpha \sim 0.15 \%$ could be achieved with {\tt SPACE}.   
In the case of the pessimistic translation to an error on $w$ considered 
in Section 4, this corresponds to $\Delta w \sim 0.6\%$; in the optimistic 
scenario, we expect a constraint of $\Delta w \sim 0.23\%$.

\section{Conclusions}

In the next five to ten years, several proposed galaxy surveys will allow high precision 
measurements of the clustering of galaxies on the scale of the acoustic oscillations at 
intermediate and high redshifts. Both photometric and spectroscopic surveys are planned,  
which will cover volumes up to tens of cubic gigaparsecs and will contain hundreds of 
thousands to hundreds of millions of galaxies. There is a clear need to ensure that theoretical 
predictions develop apace with sufficient accuracy and realism to allow such datasets 
to be fully exploited and to uncover any possible systematic errors in this 
cosmological test to uncover the nature of the dark energy. 

Early theoretical work in this area used linear perturbation theory 
(\citealt{BG2003,HuHaiman2003,GB2005}). Recently, more physical calculations have been 
carried out using N-body simulations with cubes of side 
$500-1100 \,h^{-1}\,$Mpc (\citealt{Seo2003, Seo2005, Schulz2006, Huff2006, Seo2007}).
In this paper, we have improved upon previous modelling work in three ways. 
Firstly, we have used a simulation volume comparable to the largest of the currently proposed 
spectroscopic surveys. This allows us to accurately follow the growth of density 
fluctuations on an ultra large scales in excess $100 \,h^{-1}\,$Mpc, the scales of interest 
for the acoustic oscillations, which can only be followed approximately in smaller 
computational volumes. In particular, a large volume is necessary to obtain accurate 
predictions for bulk flows, which are sensitive to the power spectrum at low wavenumbers. 
The only published work with a larger simulation volume used the Hubble Volume simulation 
(\citealt{Angulo2005, Koehler2006}). The Hubble Volume has a larger particle mass than 
the {\tt BASICC}, which restricted these studies to consider either cluster mass dark matter 
haloes (Angulo et~al. 2005) or a simple biasing scheme to add galaxies (\citealt{Koehler2006}). 
Secondly, through the use of a large number of particles, we are able to resolve the majority of the 
haloes which are likely to host the galaxies which will be observed in the forthcoming 
surveys. Thirdly, we use a semi-analytic galaxy formation model to populate the simulation 
with galaxies. Unlike other studies which use phenomenological biasing schemes or the 
halo occupation model to add galaxies, this allows us to predict the shape and amplitude of 
the galaxy power spectrum and the signal-to-noise of the clustering expected for different 
galaxy selections. 

We use our N-body simulation in combination with a galaxy formation model to make 
the connection between the linear perturbation theory prediction for the matter power 
spectrum and the power spectrum of galaxies. We do this in a series of steps, starting 
with power spectrum of 
the dark matter, looking at the impact of the nonlinear growth of fluctuations and  
peculiar motions or redshift-space distortions, before examining the power spectrum of 
dark matter haloes and, finally, galaxies. A number of conclusions are reached from this 
study: i) The nonlinear evolution of the dark matter power spectrum is apparent even on 
scales larger than the sound horizon scale. Although the deviation from linear theory 
is only a few percent, the coupled evolution of different Fourier modes means that these 
scales need to be followed accurately to get the correct behaviour at higher wavenumbers. 
ii) The form of the distortion of the power spectrum due to peculiar motions is 
extremely sensitive to the type of object under consideration, being quite different 
for the cases of dark matter, dark haloes and galaxies. Moreover, different galaxy 
selections give different redshift-space distortions. 
iii) Galaxy bias is scale dependent and sensitive to the selection applied for 
wavenumbers $k > 0.15 h {\rm Mpc}^{-1}$. 
Eisenstein et~al. (2006) discuss a technique which attempts to reconstruct 
the linear density field from an observed distribution of objects. The reconstruction 
can reduce the damping of the higher harmonic oscillations in the power spectrum, 
thereby increasing the statistical significance of the acoustic scale measurement and 
diminishing any systematic effects caused by departures from linearity.  
It will be interesting to apply this method to the galaxy samples presented in this 
paper, to see if this approach still works at the required level in the case of 
biased tracers of the linear density field.

We also present a new method to extract the dark energy equation of state parameter, 
based upon an approach put forward by \cite{Percival2007}. The method involves 
dividing the measured power spectrum by a smooth reference spectrum and comparing the 
resulting ratio to the predictions of linear perturbation theory. The algorithm has 
three key advances over earlier work, which can be credited to \cite{Eisenstein2005} and 
\cite{Percival2007}: i) The reference spectrum is derived from the measured spectrum, which 
avoids the need to apply major corrections to a linear theory reference. ii) The measured 
ratio is compared to a prediction generated using {\tt CAMB}, which is more accurate than 
assuming a parametric form for the ratio based on a Taylor expansion. iii) The linear theory ratio is modified by 
allowing the higher-order oscillations to be damped, which improves the fit to the 
measured ratio. Changing the value of the equation of state parameter is approximately 
equivalent to rescaling the wavenumber in the predicted power spectrum ratio; at $z=1$, 
a 1\% shift in wavenumber is equivalent to a $4\%$ shift in the recovered value of $w$.

We explore the constraints on the dark energy equation of state using different 
tracers of the density field. By applying our algorithm for extracting the 
oscillation scale to the {\tt L-BASICC} ensemble, we have provided the most stringent 
test to date of usefulness of baryonic acoustic oscillations for measuring the 
equation of state of the dark energy. For the case of the dark matter, there is no 
significant bias in the recovered oscillation scale, compared with the value expected 
from linear perturbation theory. Within a given simulation, we find that $1\%$ deviations 
from the underlying length scale are possible although these are only at the 1-$\sigma$ level. 
Such excursions are the result of sampling variance arising from the finite volume of 
the computational box, which are important even in a simulation of the volume of 
the {\tt BASICC}. The error on the scale factor recovered from galaxy samples is 
larger than that found for the dark matter, reflecting the lower signal-to-noise 
of the galaxy power spectrum measurements. Different galaxy selections lead to 
variations in the clustering strength and hence in the error expected in the scale factor. 

Currently, the best constraints on the equation of state parameter come either from 
combining datasets, such as the power spectrum of galaxy clustering and measurements 
of the microwave background radiation (e.g. Sanchez et~al. 2006) or from the Hubble 
diagram of Type Ia, with priors on the flatness of the Universe and the matter 
density (Riess et~al. 2004). For example, Wood-Vasey et~al. (2007) combine high 
redshift SNe Ia from the ESSENCE Supernova Survey with the measurement of the 
BAO made by Eisenstein et~al. (2005), and, assuming a flat universe, constrain 
a constant equation of state to have $w=-1.05^{+ 0.13}_{-0.12}$(stat.)$\pm 0.11$(sys.), 
consistent with a cosmological constant. Possible contributions to the systematic error 
include the degree of dust extinction in the SNe host galaxy, evolution in the properties 
of SNe with redshift and local calibration effects such as a ``Hubble bubble''. 
We have used our simulation results to forecast the accuracy with which future galaxy 
surveys will use the BAO in isolation to constrain the scale of the acoustic 
oscillations, and under certain assumptions, $w$. 
We anticipate that Pan-STARRS, with accurate photometric redshifts, 
will have an accuracy comparable to that expected for the next generation of spectroscopic 
survey (WFMOS) and could potentially reduce the statistical errors on the value of $w$ 
by a factor of 2 compared with the current constraints. However, the target of $1\%$ random errors on $w$ using BAO 
measurements is beyond the grasp of any of the surveys likely to be completed or 
even to start within the next decade.

The predictions we have presented here are idealized in a number of respects. The 
accuracy with which we expect the dark energy equation of state 
parameter will be measured assumes that the values of the other cosmological 
parameters are known with infinite accuracy. We have also neglected the impact of 
the survey window function on the power spectrum measurement; this will be particularly 
important in the case of surveys which rely on photometric redshifts. 
In future work, we plan an number of improvements: i) Use of an even larger simulation volume, 
to exceed that proposed in forthcoming surveys. One caveat on our quoted error on $w$ is that some 
of the planned surveys will be larger than the volume of the {\tt BASICC}, and will consequently 
have smaller sampling fluctuations. ii) The inclusion of the evolution of clustering along 
the line of sight. Although we have focused on $z=1$, proposed surveys will span  
a broad redshift interval centred on this value. iii) The inclusion of a survey window 
function, mimicing the angular and radial selections, and including the impact of 
errors on photometric redshifts. Such calculations represent huge challenges in computational 
cosmology, due to the volume coverage and mass resolution required in the N-body simulations 
used, and the post-processing needed to include galaxies. However, such calculations are 
essential if the BAO approach is to be used to its full potential.

\section*{Acknowledgements} 

We are indebted to Volker Springel for providing us with a stripped down version 
of his GADGET-2 code and to the Virgo Consortium for access to their supercomputing 
resources. Lydia Heck provided vital system management support which made this project 
feasible. We thank the referee, Daniel Eisenstein, for providing a thorough and 
helpful report. We acknowledge useful conversations and comments on the draft from 
Richard Bower, Shaun Cole, Martin Crocce, Ariel Sanchez, Liang Gao, Adrian Jenkins, 
Volker Springel, Enrique Gaztanaga, Francisco Castander, Pablo Fosalba and Will Sutherland.
This work was supported in part by the Particle Physics and Astronomy Research Council, 
the European Commission through the ALFA-II programme's funding of the Latin-american 
European Network for Astrophysics and Cosmology (LENAC), and the Royal Society, 
through the award of a Joint Project grant and its support of CMB. 

\bibliographystyle{mn2e}
\bibliography{bib}

\label{lastpage}

\end{document}